\newcommand{\beqn}{\begin{equation}}
\newcommand{\eeqn}{\end{equation}}
\newcommand{\beqa}{\begin{eqnarray}}
\newcommand{\eeqa}{\end{eqnarray}}

\newcommand{\xib}{\underline{\xi}}

\newcommand{\tb}{\mbox{\boldmath $t$}}

\newcommand{\Jb}{\underline{J}}

\newcommand{\gsim}{\hbox{ {\raisebox{0.06cm}{$>$} \raisebox{-0.14cm}{$\!\!\!\!\!\!\!\!\: \sim$}} } }
\newcommand{\lsim}{\hbox{ {\raisebox{0.06cm}{$<$}
\raisebox{-0.14cm}{$\!\!\!\!\!\!\!\!\: \sim$}} } }
\newcommand{\solleq}{\hbox{ {\raisebox{0.16cm}{$!$}
\raisebox{-0.04cm}{$\!\!\!\!\! =$}} } }
\documentstyle[11pt]{article}
\let\al =\alpha
\let\0=\"

\let\ka=\kappa
\newcommand{\rsim}{\hbox{ {\raisebox{0.06cm}{$>$} \raisebox{-0.14cm}
{$\!\!\!\!\!\!\!\!\: \sim$}} } }
\title{Replica Symmetry Breaking and the Kuhn-Tucker Cavity Method
in simple and multilayer Perceptrons
\thanks{based on the PhD thesis of F.~Gerl,
 Regensburg 1994}}
 \author{F. GERL$^1$ and U. KREY$^2$
\\  \\$^1$ Institut f\"ur Theoretische Physik der 
Universit\"at G\"ottingen,  
\\Bunsenstr.~9, D-30373 G\"ottingen 
\\ \\$^2$ Institut f\"ur Physik II der Universit\"at Regensburg,\\
Universit\"atsstr.~31, D-93040 F.R.G. }
       
\begin{document}

\large

\maketitle

\begin{abstract} Within a Kuhn-Tucker cavity method introduced in a
former paper, we study optimal stability learning for situations,
where in the replica formalism the replica symmetry may be broken,
namely 

(i) the case of a simple perceptron above the critical loading, and

(ii) the case of two-layer AND-perceptrons,
if one learns with maximal stability.

We find that the deviation of our cavity solution from the replica
symmetric one in these cases is a clear indication of the necessity
of replica symmetry breaking. 
In any case the cavity solution tends to underestimate the
storage capabilities of the networks.
\end{abstract}

\section{Introduction}

In a recent paper, \cite{Gerl1}, we introduced a new kind of
cavity method, with which we could solve the learning problem for
perceptrons with  $Q$- and $Q'$ state Potts model
 input and output neurons.
In this method, the  Kuhn-Tucker conditions,
which lead to optimal stability in AdaTron type learning processes,
have been built into the cavity formulation.
In subsequent papers we extended this method to the 
problem of the generalization ability
of a perceptron trained for optimal stability \cite{Gerl3}, and to the
problem of storing of correlated patterns \cite{Winkel}.

In the present paper, we apply
our method to  cases where in the replica formalism the replica
symmetry is broken, 
namely (i) to perceptrons with Ising neurons above the
critical loading and (ii) to two-layer AND perceptrons.

Cavity ideas were first applied to neural networks by M\'ezard,
\cite{Mezard1},
and Kinzel and Opper, \cite{Kinzel}. The approach of Griniasty,
\cite{Griniasty}, whose work will be discussed below in
comparison with our own findings, employs ideas introduced by
Mezard. Our formulation of the cavity method, on the other hand, was
inspired by the just-mentioned work of Kinzel and Opper. 
In another original approach, Wong \cite{Wong} employs ideas
which are related to ours and Griniasty's.

\section
{Simple perceptrons above the critical loading}

We use the definitions as in \cite{Gerl3}, which are simpler than
those we had to introduce for the Potts model case \cite{Gerl1}.  Our
perceptron has $N$ input neurons $k=1,...,N$, whose possible states
are $s_k=\pm 1$, and one output neuron $s'$, also with possible states
$\pm 1$. The couplings leading from the input neurons to the output
neuron are assumed to be real numbers, which are collected into the
coupling vector $\Jb :=(J_1,...,J_N)$ with the Euclidean length
$L:=|\Jb|=(J_1^2+...+J_N^2)^{1/2}$, which is kept fixed, while the
components $J_k$ are adapted to certain tasks by
''training processes'' (see below). The relation between input and
output is

 \beqn
 s' =\hbox{sign} \left ( \sum_{k=1}^N \, J_k\,s_k \, \right ) \quad ,
 \eeqn
 which can be considered as a binary classification of the possible
 inputs, or as {\it answers} on {\it questions}.

 Now we assume that there is a training set of $p$ input-output pairs,
 where the inputs are
 $\xib^\mu:=(\xi_1^\mu,...,\xi_N^\mu)$ for $\mu=1,...,p$, and the
corresponding desired outputs are $\zeta^\mu$. Here and in the
following, unless otherwise stated, we assume that all input
components and the outputs are independent random numbers, which take
the values $\pm 1$ with equal probability.

 The optimization task, which the perceptron then has to fulfill in
course of the training by adaptation of the components of the coupling
vector $\Jb$,
 is the minimization of a Hamiltonian, which performs a weighted count of
bad classifications (see below) of the training examples. Among
these Hamiltonians are that of Gardner and Derrida,
\cite{GardnerDerrida}, which simply counts the bad classifications,
  namely
 \beqn
 {\cal H}:=\sum_\mu V(E^\mu) := \sum_\mu \theta(\kappa -
(E^\mu/L))\,. \label{GardnerDerridaEnergie}
 \eeqn

Here $\theta (x)=1$ for $x > 0$, $=0$ else, and
$E^\mu$ is defined as usual as the ''oriented field'' 
acting on pattern $\mu$,
\beqn \label{eqn3}
E^\mu = \zeta^\mu\sum_{k=1}^N J_k \xi_k^\mu \,,
\eeqn
while $L = |\Jb |$ as above.
Bad  classifications are those, where $ E^\mu /L$ is
 $< \kappa$, i.e.~for $\kappa > 0$ they are not necessarily wrong
  but lack a prescribed amount of {\it stability}, which is measured by
            $\kappa$.

  There
are $p =:\alpha\cdot N$ such ''questions with prescribed answers'',
 i.e.~$\mu =1,...,p$, and it is assumed that the {\it loading
parameter}
$\alpha$ remains finite, while $N\to\infty$ and $p\to\infty$.
Furthermore, the  stability parameter $\kappa$
is to be maximized, if error-free classification (i.e.~${\cal H}
=0$) is possible.

Gardner and Derrida, {\cite{GardnerDerrida}}, evaluated not only the
number of errors $\cal H$ above the critical loading $\al_c(\kappa)$, where
for positive $\kappa$ error-free classification is no longer possible,
but they also tried to evaluate the so-called {\it
Almeida-Thouless}-line $\al_{AT}(\kappa)$. Above this line,
 within the replica formalism, replica symmetry breaking (RSB) is
necessary. Surprisingly, Gardner and Derrida found in
\cite{GardnerDerrida} that $\al_{AT} > \al_c$ for $\kappa
>0$: However, this was due to a subtle integration error recently
discovered by Bouten, \cite{Bouten}, who proved that
$\al_{AT}(\kappa)=\al_c(\kappa)$, as expected. Bouten also showed that
replica symmetry is always broken, if the distribution of local fields
possesses a gap. So these results, on which we will comment later,
provide a test for our cavity approach.

Other Hamiltonians, on which we comment at a later stage, are (see
\cite{GriniaGutfreund,MajerEngel}) the Perceptron function  and the
AdaTron function with 
\beqa \label{Perceptronfunction} V(E^\mu)= 
[\kappa- (E^\mu/L)]^x \cdot\theta[\kappa - (E^\mu/L)],
\eeqa 
with $x=1$ and $x=2$, respectively.

\subsection{Representation of optimal couplings}
\label{obKopplungsDarst}
Instead of minimizing the weighted error rate $f(\alpha,\kappa) :={\cal H}/p$,
 one can of course also maximize
$\kappa(\alpha,f)$ for the given training set.
 Since we are above the critical loading
 $\al_c(\kappa)$,  $f\cdot p$ of the training examples
 are badly classified.
There is however an exponentially large number $\cal N$ of partitions
of the training set into a ''good fraction'' and a ''waste
 fraction'' of size $(1-f)\cdot p$ and $f\cdot p$, respectively,
 from which one
 has to choose the optimal one. Namely, ${\cal N}\simeq \exp(c\cdot p)$,
 with $c=-f\,\ln f - (1-f)\, \ln (1-f)$.
We nevertheless assume, that every one of these combinations has
been trained for optimal stability, e.g.~with the AdaTron algorithm
\cite{Anlauf}. The optimal perceptron with error rate $f$ is then given
by the  partition leading to maximal $\kappa$.

The couplings of a perceptron trained for optimal stability can always
be expressed in the form \cite{Anlauf}
\beqn  
J_k = \frac{1}{N} \sum_{\mu \in \{(1-f)p\}} x^\mu \zeta^\mu \xi_k^\mu
\label{Jdefinition}
\eeqn
with the so-called ``embedding strengths'' $x^\mu$ of patterns, which do not
belong to the $f p$ badly classified patterns.
As can be shown using Lagrangian multipliers \cite{Anlauf,Gerl1},
these embedding strengths have to fulfill the so called Kuhn-Tucker conditions,
see below. Without restriction of generality, these are usually
formulated by fixing the length $L$ of the coupling vector $\Jb$ in
such a way that the stability limit for $\kappa > 0$ corresponds to
$E^\mu=1$, i.e.~$L=\kappa^{-1}$. With this convention, which we always
use in the following, unless otherwise stated, the Kuhn-Tucker
conditions are :
\beqn
\label{Kuhntuck}
\hbox{either} \quad (x^\mu > 0 \quad\hbox{and} \quad E^\mu =1)
\quad \hbox{or}\quad (x^\mu =0 \quad\hbox{and}\quad E^\mu >1) \: .
\eeqn
In fact, the AdaTron algorithm (without overrelaxation) 
\beqn \label{AdaTronerst}
\delta x^\mu = \hbox{max} (-x^\mu, 1-E^\mu)  \qquad 
\hbox{(sequentially or in parallel)}
\eeqn
simply fixes 
the $x^\mu$ repeatedly to values which fulfill (\ref{Kuhntuck}):
If it  converges, the conditions are necessarily obeyed. 

Using the ``oriented correlation'' matrix
\beqn
B^{\mu\nu}=\zeta^\mu\zeta^\nu 
\frac{1}{N} \sum_{k=1}^N
   \xi_k^\mu\xi_k^\nu\,.
\label{Correlatmat}
\eeqn
and the definitions (\ref{eqn3}) and(\ref{Jdefinition}),
we can write for the oriented field $E^\mu$
\beqn
E^\mu= \sum_\nu B^{\mu\nu} x^\nu \; .
\eeqn
With the Kuhn-Tucker conditions we have finally
\beqa
L^2&=& \sum_{\mu,\nu} x^\mu B^{\mu\nu} x^\nu 
= \frac{1}{N} \sum_\mu x^\mu \quad .
\label{Kopplungsnormen}
\eeqa

The basic idea of the cavity method here is, to add a pattern, i.e.\ the
``cavity'', to a set of perfectly trained patterns. By calculating
the necessary adjustments to embed this pattern we gain valuable
information about the whole system.
As in \cite{Gerl1}, we therefore add a new ''question with desired answer''
$(\xib^0,\zeta^0)$ to the training set, assuming one simple groundstate.

We note that 
  the distribution of the oriented fields $\tilde{E}^0$ acting on it,
 before any further adaptation has been performed,
which in the following always will be indicated by the \,$\tilde{}$\,, 
is Gaussian with average 0  and variance $|\Jb|^2=L^2=\kappa^{-2}$.
Of course the error rate $f$ has to remain constant. Therefore, 
as will be seen below, the
best strategy is, to give up and add the new pattern  to
   the  "waste fraction of the training set",
if the
  oriented field $\tilde{E}^0$ acting on $\xib^0$ is
 smaller than a certain number
  $Z<1$. 
   For   self-consistency, $Z$ is determined by
   \beqn
   f=\int\limits_{-\infty}^{Z}{\hbox{d}}\tilde{E}^0 \,
{\frac{\exp [ -(\tilde{E}^0)^2/2L^2]}
   {\sqrt{2\pi L^2}}} = \int\limits_{-\infty}^{z}{\hbox{d}}t \frac{\exp
   (-t^2/2)}{\sqrt{2\pi}} \,,
   \eeqn
   where $t :=\tilde{E}^0/L$ and $z :=Z/L$.
   On the other hand, if  $\tilde{E}^0$ is  $>1$, then it is
   not necessary to embed $\xib^0$ in the couplings, since it can be
   added to the set of those
    correctly classified training patterns, which need
   no explicit embedding (see \cite{Opper}) by the Adatron algorithm,
  \cite{Anlauf}. Thus, only for $Z < \tilde{E}^0 \le 1$, i.e. $z < t\le \kappa$,
  the new pattern must be embedded in the couplings, and the
  implementation strength $x^\mu$ of the other patterns must be corrected
 by $\delta x^\mu$ (see below), to compensate
  for the influence of the new pattern. 

As we have just seen, the parallel 
AdaTron algorithm would  in a first step try to embed
$\xib^0$, if necessary,  with the ''bare'' embedding $x^0 =1-\tilde{E}^0$.
This generates a perturbation  $B^{\mu 0}x^0$ of the pattern $\mu$.
In a second parallel step, all those patterns $\mu$, 
which are stored explicitly, then
have to respond by $\delta x^\mu =- B^{\mu 0}x^0$ to the disturbation by $x^0$,
because the Kuhn-Tucker conditions still have to be fulfilled.
At pattern 0 these corrections generate a response field 
\beqn \label{gBeginn}
g\, x^0 = \sum_{\mu ,( x^\mu >0)} B^{0 \mu} \delta x^\mu = 
        -\sum_{\mu, (x^\mu >0)} (B^{0 \mu})^2 x^0 \,\, , \eeqn 
which
{\it reduces} the effect of the AdaTron step with $x^0$. Therefore,
one has to enhance $x^0 =1-\tilde{E}^0$ by an
{\it amplification factor} 1/(1+g) ($>1$).
Now the $(B^{0\mu})^2$ are $1/N$ on average, see (\ref{Correlatmat}).
Therefore one gets immediately
\beqn
\sum_{\mu, (x^\mu >0)} (B^{0 \mu})^2 = \alpha\cdot P(x^\mu >0)
=: \alpha_{\hbox{\scriptsize eff}} \quad .
\label{Bmunuqu}
\eeqn 
$\alpha_{\hbox{\scriptsize eff}} $
is the percentage of exhausted degrees of freedom, i.e., if pattern 0
is as typical as the  other random patterns $\mu =1,\ldots,p$, one has to
postulate 
\beqn
g= -\alpha_{\hbox{\scriptsize eff}}  = - \alpha\cdot
P(Z<\tilde{E}^0<1) 
=-\alpha\int_z^\kappa {\cal D}t
\quad \stackrel{!}{\ge} (-1)  \quad .
\label{gistminusalphaeff}
\eeqn

Note that we have constructed our algorithm in such a way that the further
correction steps, which are necessary to regain the Kuhn-Tucker
conditions exactly, do not change the response at pattern 0 in
the thermodynamic limit when convergence is achieved. 
A proof is given in the Appendix, where we
also show that in this limit one can assume that the
$x^\mu$ and $B^{\mu\nu}$ are statistically independent.

With our approach, the final embedding strength for $\xib^0$ is then
given by $(1- \tilde{E}^0)/(1+g)$. Furthermore, we now identify the
distribution of embedding strengths of all patterns with that of
pattern $\xib^0$.  Putting the Gaussian distribution of the cavity field $
\tilde{E}^0$, felt before training, into (\ref{Kopplungsnormen}) gives
\beqn L^2 = \alpha L \int_z^\kappa {\cal D} t \, \frac{\kappa -
t}{1+g} \; .\eeqn 
 With $L=\kappa^{-1}$ this implies 
\beqn 1 =
\frac{\alpha}{1+g} \int\limits_z^\kappa {\cal D}t \, (\kappa-t) \kappa
\; .  \label{Wongvgl} \eeqn 
After multiplication with $1+g$, this
finally leads to our ''Kuhn-Tucker cavity result''
\beqa 1 &=&\al \int\limits_z^\ka {\cal D}t +\al\ka
\int\limits_z^\ka {\cal D}t\,(\ka-t) \label{KTresult} \\
 &=& \al\,\, (1+\ka^2)\int\limits_z^\ka{\cal D}t +
 \frac{\alpha \ka }{\sqrt{2\pi}}
(e^{-\ka^2/2}-e^{-z^2/2})\,.
\label{KTresult2}
\eeqa

This result will now be compared with that of the cavity approaches of
Griniasty, \cite{Griniasty}, and Wong \cite{Wong}. Their different
result is
equivalent to the replica formalism {\it in the replica-symmetric
approximation}, \cite{GardnerDerrida,GriniaGutfreund}; therefore we use
the suffix ''RS'' to indicate their result. We have already shown
in [1] that below $\al_c$, where replica-symmetry is exact, our
Kuhn-Tucker cavity approach and the RS approach agree; however in the
present situation, where $\al$ is $>\al_c$, they disagree.

When Griniasty derives the constant of integration in \cite{Griniasty},
he uses a simple method which in all cases known to us gives the
correct RS result: The reaction factor $g$ is assumed to vanish, while
at the same time also the $B^{\mu \nu}$ with $\mu\ne\nu$ are
neglected.  With these two neglections one obtains instead of
eqn.~(\ref{Kopplungsnormen})
 \beqa L^2&=&\frac{1}{N} \,\vec{x}^{\,
T}_{\hbox{\tiny RS}} \! \stackrel{\leftrightarrow}{B}_{\hbox{\tiny
RS}} \vec{x}_{\hbox{\tiny RS}} \:=\: \frac{1}{N} \sum_\mu (
x^\mu_{\hbox{\tiny RS}} )^2\quad . \eeqa 
Thus,
instead of eqs.~(\ref{KTresult}) and (\ref{KTresult2}),
 the ''RS'' cavity result would be 
 \beqa 1&=&
\alpha_{\hbox{\tiny RS}} \int\limits_z^\kappa {\cal D}t \:
 (\kappa -t )^2 \nonumber \\
&=&  \alpha_{\hbox{\tiny RS}} (1+\kappa^2)  \int\limits_z^\kappa {\cal D} t 
 \:+ \: \frac{\alpha_{\hbox{\tiny RS}} \kappa}{\sqrt{2 \pi }} 
( e^{- \kappa^2/2}  - e^{-z^2/2})
+ \underbrace{ \frac{\alpha_{\hbox{\tiny RS}}}{\sqrt{2 \pi }} 
(z-\kappa) e^{-z^2/2}}_{=: M}   
\label{obRSausger}   \quad. \qquad
\eeqa
 Eqn.~(\ref{obRSausger}), which is identical with the result obtained by
Griniasty \cite{Griniasty} or Wong \cite{Wong}, agrees with the result
of the replica calculation of Gardner and Derrida
\cite{GardnerDerrida,GriniaGutfreund} from the replica-symmetric
approximation; the expression abbreviated by
 $M$ in eqn.~(\ref{obRSausger}) yields
 the difference between our $\alpha_{\hbox{\tiny Cav}}$ (=$\alpha$
obtained from eqn.~(\ref{KTresult2}))
and  $\alpha_{\hbox{\tiny RS}}$. Because of $z < \kappa$, it is
$M\le0$ and therefore
 $\alpha_{\hbox{\tiny RS}} \ge \alpha_{\hbox{\tiny Cav}}$.
For $z \to -\infty$, i.e.~for $f \to 0$, one has $M=0$, and
thus for $\al \le\al_c$ it is
    $\alpha_{\hbox{\tiny RS}} = \alpha_{\hbox{\tiny Cav}}$, as already
mentioned.

Although the numerical results differ, there is an interesting formal
relationship between the basic equations in Wong's approach \cite{Wong} 
and the one presented here: Eqn.~(6) in \cite{Wong}
agrees formally with our eqn.~(\ref{Wongvgl}), and our reaction
strength $g$ corresponds to $-\alpha\cdot\chi$ in \cite{Wong}.
However $-\alpha\cdot \chi$ in \cite{Wong} differs from our
$g$, which is given by eqn.~(\ref{gistminusalphaeff}), by a term
corresponding just to the expression $M$ in eqn.~(\ref{obRSausger}).
The difference arises from a $\delta$-function contribution to
$\lambda'(t)$ at $t=z$ in \cite{Wong},
where $\lambda(t)$ and $t$ in \cite{Wong} are the
oriented fields {\it after} training and {\it before} training,
respectively.  Probably this difference, which only comes into play
above $\alpha_c$, where $M$ is $\ne 0$, is relevant with respect to
the combinatorial explosion, which conflicts with the assumption of a
unique optimum made in all the approaches. Research on this problem is
in progress.

In Fig.~1, for error rates of $f=0.2$ and 0.02, the
learning capacities $\al(\kappa ,f)=\alpha_{\hbox{\tiny Cav}}$ and
$\alpha_{\hbox{\tiny RS}}$, respectively,
as  obtained from  our Kuhn-Tucker cavity
 theory, eqn.~(\ref{KTresult}),
 and  with the replica-symmetric
 approximation  (\ref{obRSausger}) respectively, are
 plotted against the stability $\kappa$.
Obviously, the learning capacity obtained with the Kuhn-Tucker cavity
theory is lower, particularly at small values of $\kappa$, i.e.~for $\kappa
=0$ and $f=0.2$, $\alpha_{\hbox{\tiny Cav}}$ is only 10/3, whereas
$\alpha_{\hbox{\tiny RS}}$ is 15.53. This means at the same time
that for given $\kappa$ and
$\alpha$, our error rate $f$ would be larger than that obtained with
eqn.~(\ref{obRSausger}). This is a strong hint that 
the assumption of a unique optimum for the distribution of those
patterns, which are put to waste, is wrong.  Therefore, one cannot say in
advance, which approach is better: Rather what has been gained is the
following: Since both approaches agree
below $\al_c(\kappa ,f=0)$, but not for $f >0$, we can use the
different results as a criterion
for the necessity of replica-symmetry breaking
for $f >0$, which agrees with the recent rigorous proof
of $\cite{Bouten}$.

\subsection{Comparison with  a One-Step-RSB calculation}
\label{RSBMajKap}

Majer {\it et al.}, \cite{MajerEngel},
 have performed a calculation above $\al_c(\kappa, f=0)$ within a one-step
replica-symmetry-breaking approximation.
 From  eqn.~(\ref{GardnerDerridaEnergie}), they obtained
the following result for the free  energy
 $\langle f_{\hbox{\tiny E}} \rangle= \alpha f(\alpha ,\kappa)$:
\beqa 
- \langle f_{\hbox{\tiny E}} \rangle
 &=& \min_{x, q_0, w} \bigg\{ \frac{q_0}{2x(1+w \Delta q)}
 \:+\: \frac {\log (1+w \Delta q)}{2 w x}  \nonumber    \\
&&\;+ \:\frac{\alpha}{wx} \int {\cal D} z_0 \:\ln \bigg[ 
\int^{\frac{A}{\sqrt{\Delta q}}}_{\frac{A-2x}{\sqrt{\Delta q}}}
 {\cal D}z_1
\exp\Big\{-\frac{w}{2} \big(A- z_1 \sqrt{\Delta q}\: \big )^2 \Big\}  \quad
\nonumber\\ \label{Parameterbest}
&&\;+\:\exp\{-wx\} \Phi\big(\frac{A-2x}{\sqrt{\Delta q}}\big)
\:+\: \Phi \big( - \frac{A}{\sqrt{\Delta q}}\big)\: \bigg] \bigg\} \; ,
\eeqa
with $A= \kappa- z_0 \sqrt{q_0}$ and $\Delta q = 1-q_0$.
The parameters $q_0,\, w$ and $x$ have to be chosen such that
the number of errors is maximized. In the limit $q_0\to 0$ one
regains the RS result.

\label{Stabvf}

In  Fig.~2 the stability $\kappa$ is plotted against the
error rate $f$ for three different values of $\alpha$.
Obviously there is only a small difference between the results
of the RS and the 1-step RSB calculation, in contrast to our
cavity results, which differ considerably from both replica calculations, 
i.e.~with our estimate, 
the stability increases much more slowly with increasing $\al > \al_c$.

For $f \ll 1$, we can compare these different estimates with simple
simulations: For $\alpha<2$ one can train perceptrons to optimal
stability by means of the AdaTron algorithm. If one then skips the
pattern with the largest embedding strength, i.e.~the one which was
most difficult to store, and re-learns the remaining patterns, one
gets an enhanced stability, which
 agrees within the error limit with the replica calculation, and
  not with the cavity results, as we found in the simulations.

Thus, for $f >0$, our Kuhn-Tucker cavity method yields a 
non-sufficient approximation
for $\kappa(\alpha,f)$. In the following we try to find the reasons
for the discrepancy and to estimate at the same time the  quality of the
1-step RSB calculation. For this purpose, we need the distribution of
the oriented fields $t$, which according to
Majer et al., \cite{MajerEngel}, is
\beqn
\varrho(t)\:=\: \int {\cal D}z_0  \:
\frac{
\int {\cal D}z_1 \: \exp \left\{ -wx \Big ( V(\lambda_0) +
\frac{(\lambda_0- z_0 \sqrt{q_0} - z_1 \sqrt{\Delta q } )^2 }{2x}
\Big  )\right \} 
\delta (t- \lambda_0) }
{
{\int\cal D}z_1 \: \exp \left\{ -wx \Big (V(\lambda_0) +
 \frac{(\lambda_0- z_0 \sqrt{q_0} - z_1 \sqrt{\Delta q })^2}{2x}
 \Big )\right\} 
}\,.
\label{MajerFeld}
\eeqn 
where $V$ has been defined for the  Gardner-Derrida rule in
 (\ref{GardnerDerridaEnergie}),
and where $\lambda_0$ has to be chosen such that the exponent is minimized.

The parameters $q_0,\,w$ and $x$ are determined for given $\al$ and $\ka$
from eqn.~(\ref{Parameterbest}). With eqn.~(\ref{MajerFeld})
the probability  distribution of the local fields can be determined;
in particular  a $\delta$-peak at $t=1$ is obtained, which
represents the patterns, which are embedded explicitly by the training.
However, our interest is in the field-distribution {\it before} the additional
training.

This distribution can be obtained by extending Griniasty's interpretation
from the RS results in \cite{Griniasty} to the present case.

 According to \cite{Griniasty}, after addition of the pattern $\xib^0$, the
 quantities
  $ z_0 \sqrt{q_0} - z_1 \sqrt{\Delta q }$ and $\lambda_0$ are
   the local oriented fields felt
  before and after the additional training, respectively.

 The exact value of  $\lambda_0$ results from a compromise between
 the increase of the energy  $V(\lambda_0)$ for patterns, which are
 badly classified by the perceptron, and the term
$(\lambda_0- z_0 \sqrt{q_0} - z_1 \sqrt{\Delta q })^2/2x$ representing
the increase in energy of the  $(1-f)p$ patterns, which had been stored before
the addition of the new pattern.

Thus we can determine the field before the corrections by simply eliminating
  $V(\lambda_0)$ and $\lambda_0$ in
(\ref{MajerFeld}): 
 With the abbreviations
 $u_0= z_0 \sqrt{q_0}$ and $u_1= z_1 \sqrt{\Delta q}$,
  the integrand in eqn.~(\ref{MajerFeld})  is
\beqa
&&\frac{1}{\cal N} \int {\cal D}z_1 \bigg[ \exp \left \{
-\frac{w}{2} (\kappa - u_0- u_1 )^2 \right\}
\theta(\kappa - u_0- u_1 ) 
\theta( u_0 + u_1 -(\kappa-\sqrt{2x}))  \nonumber\\
&& \;+ \: \exp(-wx) \,\theta(\kappa-\sqrt{2x}- (u_0 + u_1))
\:+\: \theta(u_0 + u_1 - \kappa)  \bigg ] 
\delta(t-(u_0 + u_1 )) \nonumber \\ 
&=&  \frac{1}{\cal N}
\exp \left\{ \frac{(t - u_0)^2}{2 \Delta q}\right \}  \bigg [
\exp \{ -\frac{w}{2} (\kappa-t)^2\} \theta(\kappa-t) \theta(t-(\kappa-\sqrt{2x})) 
\nonumber\\ &&
\qquad \qquad \;+\: \exp\{-wx\} \theta(\kappa-\sqrt{2x} -t)
\:+\: \theta(t-\kappa) \bigg ]  \quad . \label{Feldintegr}
\eeqa

Here the first term describes the patterns, which have been successfully
 embedded (i.e.~with $V=0$),
the second term those,  which are badly classified and put to the
waste
 (i.e. with $V=1$ and $\lambda_0-z_0\sqrt{q_0}-z_1\sqrt{q_1}=0$),
and the final term the patterns which are stored without embedding.
The denominator ${\cal N}$  in (\ref{Feldintegr})  is the
integral over all fields $t$ in (\ref{Feldintegr}) and serves for
normalization.
Thus, we have no longer a Gaussian field-distribution before learning.

In the cavity picture, this non-Gaussian distribution of the fields
acting on an untrained
pattern stems from the fact that there are now many ground states
available.  If we add a pattern to such a multitude, every groundstate
will still see a Gaussian distribution of the local field
$\tilde{E}^0$.  However the particular ground state, which after training
appears as the one with the highest stability, is likely to have a
higher-than-average local field for the new pattern, thus being able
to store it more easily. Our field-distribution before learning is
then effected by this selection.  At the end of this section we will
shortly comment on how an intrinsic cavity approach for RSB should
take this selection effect into account.

\label{FeldvertRSB} 

In Fig.~3, for $\kappa=1$ and $\al=1$, the local fields are presented
for the three approximations mentioned, namely for our cavity method,
and for the RS and RSB1 approximations.
One can see that in RSB1, the distribution of the local fields before
learning the additional pattern, although strongly non-Gaussian,
 is still everywhere continuous, and for $ t =\kappa$ even continuously
 differentiable.

Comparing the results of our cavity theory with the RS approximation,
 one can see that for increasing $\al > \al_c$ 
 \begin{itemize}
 \item on the one hand,  the {\it error rate} $f$
  obtained with the  RS theory is much
 better than that obtained with our cavity method
 (if the error rate obtained with RSB1 is
considered as target approximation), see Fig.~2;\,\, whereas
\item on the other hand, the value of the
 limiting negative field value $t:=t_0=z/\ka$
($\approx -0.5$ in Fig.3), below which patterns are no longer learnable,
is almost exactly the same with our simple cavity approach and the much
more complicated RSB1 calculation.
\end{itemize}

If we accept the local fields from RSB1 as a good approximation, then 
again the Kuhn-Tucker conditions (\ref{Kuhntuck})
must be fulfilled after learning, and with  (\ref{KTresult})  we can
calculate the loading $\alpha_{\hbox{\tiny RSB-KT}}$
from the RSB1 field-distribution, calculated with the parameters
$q,w$ and $x$. To this purpose, in the above-mentioned formula one only has
to replace the Gaussian measure ${\cal D}t$ by the  RSB1 field-distribution of
those patterns, which are expicitly embedded, i.e.~from $t_0$ to $\kappa (=1)$
in Fig.~3. This is related to $\alpha_{\hbox{\tiny RSB1}}$
in a similar way as  $\alpha_{\hbox{\tiny Cav}}$
is related to $\alpha_{\hbox{\tiny RS}}$.

For three values of $\ka$, the result 
$f_{\hbox{\tiny KT}}:=f_{\hbox{\tiny RSB1-KT}}$
 of such a calculation is presented
in Fig.~4.
For comparison also the results of the two simple approximations,
$f_{\hbox{\tiny Cav}}$ 
 and $f_{\hbox{\tiny RS}}$,
 are presented, together with the RSB1 replica
result $f_{\hbox{\tiny RSB1}}$. Obviously, the improved cavity result
$f_{\hbox{\tiny KT}}$ is only
slightly higher than  $f_{\hbox{\tiny RSB1}}$, which is already
a criterion for the
quality of the RSB1 calculation compared with the RS theory.
 
One expects therefore that the rigorous
result should lie between our $f_{\hbox{\tiny KT}}$ as an upper and
 $f_{\hbox{\tiny RSB1}}$ as a lower bound, 
so that further replica symmetry breaking steps should give only a
slight improvement. Precisely, we expect both a slow monotoneous
 {\it increase} of the error
rate $f_{\hbox{\tiny RSBn}}$  with increasing $n$ in a RSBn calculation,
\cite{FontanariTheumann}, and a slow monotoneous {\it decrease}
of $f_{\hbox{\tiny RSBn-KT}}$,
and at the same time a decrease of the relative number of explicitly
embedded patterns and of the averaged embedding strength.
 As a consequence, $\alpha_{\hbox{\tiny RSBn-KT}}$, which is
 calculated from a formula analogous to (\ref{KTresult}), increases
 slowly. For $n\to\infty$,
the RSBn replica result and the RSBn-KT cavity
result derived from the RSBn field-distribution
should agree.

We have thus found a method leading to an upper bound for the error rate
$f(\alpha ,\kappa )$ as a function of $\al$ and $\ka$, 
if the local fields before
re-learning are known. In \cite{FontanariTheumann}, Fontanari and
Theumann derive an upper bound for $f$ at $\ka$=0, evaluated in RS
approximation at the AT line for finite temperatures: For $\al\lsim 50$,
our RSB1-KT upper bound is lower, whereas for $\al \rsim 50$ the bound in
Fig.~2 of \cite{FontanariTheumann} is lower.

That the RSB1 itself is not yet sufficient, has already been 
shown by  Majer and Engel
 \cite{MajerEngel}. Moreover, also the fact that after our RSB1-KT cavity
 learning there is still a {\it gap} in the field-distribution, which is
 according to Bouten  \cite{Bouten} responsible for RSB, points to this fact.
Additionally, in this approximation the condition is violated that the
number of explicitly embedded
patterns should not exceed the number  $N$ of coupling degrees of
freedom, see eqn.~(\ref{Feldintegr}).

In connection with Wong's paper, \cite{Wong}, it is interesting to
note the following: Using the RSB1
distribution of fields in connection with  eqn.~(6) in \cite{Wong}
one reproduces self-consistently the RSB1 result for $\alpha$ of
Majer and Engel, \cite{MajerEngel}.  The slight difference to our
result arises again from a contribution of a $\delta$-peak, which is
present in Wong's approach, but not in our's. This contribution, much
smaller now,
which in turn originates from the mentioned gap, leads again to terms
corresponding formally to $M$ in eqn.~(20). Thus once more the difference
between our results and those obtained with Wong's approach, both
starting with the above-mentioned RSB1 field distribution, shows that
also the RSB1 calculation, albeit a good approximation, is not yet exact.  

The results discussed here have recently been supported by a
complicated 2-step-RSB calculation: 
Whyte and Sherrington find in  \cite{Sherrington} 
that in fact the next step in the 
replica breaking scheme raises the error rate $f$, but only by an amount
which is typically ${\cal{O}}(10^{-4})$. This agrees very well with the
predictions that we could make from our findings.

As already mentioned, the results of the cavity methods are
necessarily insufficient above $\al_c$, since the combinatorial
possibilities to select the "waste patterns" are not considered.
However, with our approach it should also be possible to get
equations, which are equivalent to RSB1, without using the replica
trick, i.e.~as in the seminal book \cite{Mezard} on spin glasses.

To achieve this, one considers a multitude of ground states, which are
ordered in an ultrametric structure. If one decreases the stability
constraints, the number of different ground states is assumed to
increase exponentially. If one now adds a pattern to this ensemble, a
lower embedding strength is therefore favoured, which gives a higher
storage capacity compared to the Gaussian one of our
eqn.~(\ref{KTresult}). 
This approach has the appealing trait, that the number of patterns,
which have to be misclassified, gets lower
as more degrees of freedom become available for the approximation.
Within the replica method, in contrast, one has to take the ``worst''
value of the order parameters for technical reasons. 
Work on an intrinsic cavity approach for
$\al > \al_c$ is in progress.

\section{The AND-machine}
\label{AndMachine}

For multilayer perceptrons, replica symmetry breaking is a general
phenomenon, when optimal learning capacity for a given stability
or optimal stability for given capacity are required.
 As a consequence, the optimal capacity is not easily estimated.
Griniasty and Grossman \cite{GriniaGross} have calculated
the storage capacity of the AND-machine and gave arguments,
which made a suppression of replica symmetry breaking credible.
However, with our method we are able to
check their proposal and find that replica symmetry is broken.

\subsection{Model description}

The AND-{\it machine} is a simple example for a {\it multi-layer}
perceptron. Multilayer perceptrons have been introduced to overcome
the limitations of single layer perceptrons, which are limited to {\it
linearly separable classifications} \cite{MinskyPapert}, whereas with
sufficiently many units in the additional layer(s) between the input and
output units, every Boolean function can be implemented.  However, at
the same time, the analytical treatment of the models becomes much more
complicated, and in general, the space of solutions is no longer
simply connected.  The RS approximation yields then results which
contradict the exact bounds derived by Mitchison and Durbin
 \cite{Mitchison}. However, the necessary RSB calculation is rather
complicated for such models. A multilayer model, which has been
studied in this way, is the so-called
 {\it committee
machine} with 3  (or any other odd number $N_h$ of) {\it hidden units}
\cite{BarkaiHansel,EngelMehrschicht}. Here the intermediate {\it
hidden
layer} consists of three neurons, and the output is given by the
{\it majority vote} of these
 {\it hidden neurons}. The maximal capacity of the system
decreases from $\alpha_c \simeq 4.02$ for the RS approximation to
$\alpha_c \simeq 3.0$ for the RSB-{\it ansatz}. This is for the case
of non-overlapping receptive fields, see fig.~5.

In contrast to the mentioned case of the 
{\it committee} machine, for the AND machine the number
$N_h$ of intermediate neurons is arbitrary ($\ge 2$); here the output
unit gives a positive vote iff all intermediate neurons vote with
+1. This AND machine was at first treated by Griniasty and Grossman,
\cite {GriniaGross},
both with non-overlapping and also with overlapping receptive fields
(NRF and ORF cases, respectively). Within a replica-symmetric approach,
 the authors find for
the case of an equal number of patterns with output $+1$ and $-1$ for
$N=2$ intermediate neurons a critical capacity of $\al_c \simeq 3.5$
($\al_c\simeq 3.3$ with simulations), for the case of overlapping
receptive fields, and $\al_c\simeq 3.66$ for the NRF case,
\cite{GriniaGross}.  At the end of their paper, Griniasty und
Grossman 
 discuss the validity of their RS
approximation and find support for their assumption that only a single
minimum contributes to their solution.  Griniasty \cite{Griniasty}
repeated the calculation with his cavity method and got the same
results. Wong's recent result \cite{Wong} on this problem will be
commented upon below.  
In the following, we use our different cavity method.

\subsection{Calculation of the learning capacity}

As  we have seen, our method yields a convenient way to recognize the
necessity of RSB. 
Additionally one gets a rough quantitative estimate,
to which extent the actual solution is approximated.
At first we study the  AND machine with $N_h=2$ and non-overlapping receptive
fields (NRF case).
Since this does not necessitate an additional effort and the
argumentation is simplified, we assume that both subnetworks are
trained with optimal stability, as long as the optimal capacity
$\alpha_c$ is not yet reached.

We assume  $p_+= \alpha_+ N$ and $p_-=\alpha_-N$
patterns with positive resp.~negative output.
Then the parameter $b$ is defined via
\beqn
\alpha_\pm\:= \: \frac{1 \pm b}{2} \:\alpha \quad .
\label{ANDbias}
\eeqn
The $(+)$-patterns must be trained in both subnets, since both
intermediate
 neurons must vote positively, whereas for
the $(-)$-patterns one can choose at will a subnet with a negative vote.
From this fact, Griniasty and Grossman \cite{GriniaGross}
derive the sharp bounds
\beqn
\frac{6}{2+b} \: \le\: \alpha_c\: \le\: \frac{4}{1+b}
\label{ANDbound}
\eeqn 
for the maximal capacity $\alpha_c$, which also applies with
overlapping receptive fields (ORF case). Moreover, recently Wendemuth
\cite{Wendemuth}
was able to generalize the method of Mitchison and Durbin
\cite{Mitchison}
and obtained for the  AND-machine as a lower bound the sharper condition
\beqn
\frac{8}{4-(1-b) \log_2(3) } \ge \alpha_c \quad ,
\label{Wendbed}
\eeqn
which also applies both to the NRF and the ORF case.

Let us consider fictitiously all possibilities to distribute the
responsibility for the $(-)$-patterns among the $N_h$ subnetworks;
these patterns, including the $(+)$-patterns, shall then be trained to
optimal stability. Afterwards we select that distribution, which leads
to maximal stability $\kappa=\min \{\kappa_1,\kappa_2\}$.

In both subnetworks $i=1,2$ we have $N$ input
neurons, and in both subnetworks the couplings are defined with
embedding strengths $x^\mu_i$ as

 \beqn
J_{ik} \: = \: \frac{1}{N} \sum_\mu x_i^\mu \zeta^\mu \xi_{ik}^\mu \quad
.  \label{JANDKopplungen} \eeqn

Here $\zeta^\mu$ is the desired final output. As before, we define the
oriented correlation matrix
$\stackrel{\leftrightarrow}{B}$, the length $L$ of the coupling vectors
 and the oriented local fields \,\,\,$\vec{E}$ 
\beqa
B^{\mu\nu}_i & := & \frac{1}{N}\: \zeta^\mu\zeta^\nu
\sum_k \xi_{ik}^\mu \xi_{ik}^\nu \\
L^2_i &:=& \underline{J}^T_i \underline{J}_i \:=\:
\frac{1}{N} \sum_{\mu,\nu} x_i^\mu B^{\mu\nu}_i x_i^\nu \\ 
E^\mu_i &:=& \sum_\nu B^{\mu\nu}_i x_i^\nu  \quad .
\eeqa
The embedding strengths $x^\mu_i$, for $i=1,2$ and all $\mu$,
  together with the $E_i^{\mu}$, fulfill again the KT conditions
\beqn
\hbox{either} \quad  (x^\mu_i >0 \;\; \hbox{and} \;\; E^\mu_i=1)
\quad \hbox{or} \quad  (x^\mu_i=0  \;\; \hbox{and} \;\;  E^\mu_i>1)
 \quad \;\; .  \label{ANDFix}
\eeqn
  Additionally, we know that for our optimal choice, a
$(-)$-pattern can have a positive embedding strength only at one of
the subnets, since otherwise the less stable subnet could enhance the stability
by reducing its unnecessarily positive
 $x^\mu_i$ to 0 and relearning the other patterns. Furthermore, in
the thermodynamic limit, both stabilities should be equal, 
i.e.~$L^2_1= L^2_2=: L^2$.
Let us assume again that there is only one groundstate, and add a new
pattern. 
 Again this feels a normally distributed random oriented field
 $\widetilde{E}^0_i$ with variance  $L^2$ in each subnet. A
$(+)$-pattern must be classified correctly in both subnets. So  we
need, in case of  $\widetilde{E}^0_i<1$, positive embedding strengths
 $x^0_i= (1-\widetilde{E}^0_i)/(1+g_i)$, where the response factors
$g_i$ have still to be determined. In contrast, for a $(-)$-pattern we
can choose the subnet with negative intermediate output.
 As we will see below, it is
best to embed the $(-)$-pattern in that subnet, where the oriented
field is larger, so that the embedding strength can be smaller.
 The field-distribution  $P(t)$  for the larger oriented
field $t$, with normalized couplings,
 ($\widetilde{E}^0_{\hbox{\scriptsize max}}= Lt$), can be calculated
from the Gaussian distribution of the fields $(t_1,t_2)$ of the two
 sublattices as follows:

\beqa P(t)&=&2\,P(t_1=t, t_2<t) \: = \;
        \frac{2}{\sqrt{2\pi}} \:e^{-t^2/2} \int\limits_ {-\infty}^t
{\cal D}t_2 \nonumber\\ \label{PvontNRF} &=&
 \frac{2} {\sqrt{2\pi}}\: e^{-t^2/2}
\:\Phi(t) \quad .  \eeqa
 If again we assume that $x^\mu$ and $B^{\mu\nu}$ are uncorrelated,
then the answer $g$ of the patterns, which had already been
implemented and now must keep the KT conditions, is similar to
eqn.~(\ref{gistminusalphaeff}), namely

\beqn
\: -\sum_{\mu \atop (x^\mu >0) } (B^{0\mu})^2 \:=\: 
-\alpha_{\hbox{\scriptsize eff}} \: = \: - \alpha P(x^\mu>0 ) \quad. 
\label{AntAND}
\eeqn

Identifying again the distribution of the embedding strengths with the
probability distribution for $x^0$ and normalizing the couplings to 1,
we get
\beqn
g \:=\; - \alpha_+ \int\limits_{-\infty}^\kappa {\cal D}t
 \: -\: \:\alpha_-  \int\limits_{-\infty}^\kappa {\cal D}t \,\Phi(t)
\label{gvonAND}
\quad .
\eeqn

Here the first and second parts describe the influence of $(+)$- and 
 $(-)$-patterns, whereby compared with eqn.~(\ref{PvontNRF}) a factor
1/2 was taken into account, since only one of the two subnets
ist needed for $(-)$-patterns.
 The capacity for finite stabilities is again
calculated from the KT conditions (\ref{ANDFix}) through 
\beqa
 L^2_i &=& \frac{1}{N} \sum_{\mu\nu} x^{\mu }_i B^{\mu \nu}_i x^\nu_i
 \: = \: 
\frac{1}{N} \sum_\mu x^\mu_i E^\mu_i 
= \frac{1}{N} \sum_\mu x^\mu_i  \\
 &=& \alpha \int \hbox{d}x_i \:x_i
w(x_i) 
= \frac{L_i }{1+g} 
\int\limits_{-\infty}^\kappa {\cal D} t  \,
 ( \alpha_+ + \alpha_- \Phi(t) )\cdot
(\kappa-t )\,.  \quad 
\label{KuhnTuckAND}
\eeqa
Again, with $\kappa = 1/L$ and multiplying by $(1+g)$, we obtain finally
 from our cavity method
\beqa
1 &=& \alpha \int\limits_{-\infty}^\kappa {\cal D} t  \:
\left(
\frac{\raisebox{-0.03cm}{\hbox{\footnotesize 1 + {\it b}}} } 
 {\raisebox{0.03cm}{\hbox{\footnotesize 2}} }   \: +\: 
\frac{\raisebox{-0.03cm}{\hbox{\footnotesize 1 -- {\it b}}} } 
 {\raisebox{0.03cm}{\hbox{\footnotesize 2}} } 
\,\Phi(t) \right) \cdot (1 + \kappa( \kappa -t))
\quad .
\label{CavANDErg}         
\eeqa 

Obviously, every other strategy to store a pattern would lead to a
smaller learning capacity with our method. The maximal capacity for
$\kappa = 0$ is then with our  method (i.e.~with $\alpha =
\al_{\hbox{\tiny Cav}}$)

 \beqa
\label{eqn38}
 (\alpha_c^{-1})_{\hbox{\tiny Cav}} &= & \frac{1+b}{2}
\int\limits_{-\infty}^0 {\cal D}t
 \;+\; \frac{1-b}{2} \int\limits_{-\infty}^0 {\cal D}t \,\Phi(t)
\nonumber \\
&=& \frac{1+b}{4} \:+\: \frac{1-b}{16} \; = \; \frac{5+3b}{16} \quad .
\eeqa

This corresponds again to the limit $g=-1$ of eqn.~(\ref{gvonAND}) and is
compatible with the exact bounds (\ref{ANDbound}), although it is
somewhat lower than the improved lower bound (\ref{Wendbed}).
 This, again, is no surprise, since we always expect a somewhat too
 low estimate from our method in case of RSB situations.
  
Now, in contrast, we perform the ''handwaving approach'' mentioned by
Griniasty, see \cite{Griniasty},
to neglect the non-diagonal terms of $\stackrel{\leftrightarrow}{B}$ 
and assuming at the same time $g=0$.  
Instead of eqn.~(\ref{CavANDErg}) 
this leads to ''RS'' results, namely 
 \beqa 1&=&\frac{1}{N} \,\vec{x}^{\,
T}_{\hbox{\tiny RS, i}} \! \stackrel{\leftrightarrow}{B}_{\hbox{ \tiny
RS, i}} \vec{x}_{\hbox{\tiny RS, i}} \:=\: \frac{1}{N} \sum_\mu (
x^\mu_{\hbox{\tiny  RS, i}} )^2 \nonumber
 \\ &=& \alpha_{\hbox{\tiny RS}}
\int\limits_{-\infty}^\kappa {\cal D} t \: \left(
\frac{\raisebox{-0.03cm}{\hbox{\footnotesize 1 + {\it b}}} }
 {\raisebox{0.03cm}{\hbox{\footnotesize 2}} }   \: +\: 
\frac{\raisebox{-0.03cm}{\hbox{\footnotesize 1 -- {\it b}}} } 
 {\raisebox{0.03cm}{\hbox{\footnotesize 2}} } 
\,\Phi(t) \right)
 \cdot ( \kappa -t)^2 \quad . \label{RSANDEnd} \eeqa 
 For $\kappa=0$ this yields instead of eqn.~(\ref{eqn38})
the result of Griniasty and Grossman, \cite{GriniaGross}, namely 
\beqa
(\alpha_c^{-1})_{\hbox{\tiny RS}}
 \;=\; \frac{1+b}{4} \:+\: \frac{1-b}{2} \: 0.045422528\ldots
\quad . 
\eeqa

These results are for non-overlapping receptive fields (NRF).
For finite $\kappa$ and $b=0$, the 
maximal capacity $\alpha_c$ is compared for both
approximations in fig.\ 6. Additionally, also the result for
overlapping receptive fields (ORF,
$\alpha_{\rm full}$, see below) is presented as the
dashed curve, which is slightly lower for the present case, but not
always. This will be discussed later in more detail.

The fact that the RS result differs from our cavity result is,
according to our experience, a strong hint for the necessity of RSB.
The extent of replica symmetry breaking seems to increase with higher
loading and larger percentage of 
 $(-)$-patterns.

For $\al_+=\al_-$ the maximal capacity according to our theory is
$(\alpha_c)_{\hbox{\tiny Cav}}=3.2$,
 whereas eqn.~(\ref{Wendbed}) leads to $\alpha_c\ge 3.31$ and the RS approximation
 to $(\alpha_c)_{\hbox{\tiny RS}}=3.667$. Probably
the true result is again smaller than the RS value, but not too far from it.
Thus, for $\alpha_+=\alpha_-$, the RS appoach yields again a good estimate
although replica symmetry is broken:
The limit $g=-1$, above which the system is obviously over-determined
concerning the number of couplings, is already reached  at $(\alpha
_c)_{\hbox{\tiny Cav}}$, i.e.\ below  $(\alpha_c)_{\hbox{\tiny RS}}$.

The local stability has not been checked by replica calculations.
Recently, however, Wong could use his cavity method \cite{Wong} 
to check a large class of multilayer perceptrons and found
that for the AND-machine replica symmetry indeed is broken.
 
Our cavity method is not only applicable to the present case of an  NRF-AND
machine, but also for NRF-machines with {\it arbitrary Boolean output}
functions:
If one has found the optimal strategy similar as above, the steps
leading to eqn.~(\ref{CavANDErg}) are identical, and one only has
to substitute the embedding strengths, which compensate  the
normally-distributed random field, in this equation.

Formally this leads to the following equation for the storage capacity
$\alpha= \min_i\{\alpha_i\}$:
If $\tb$ is the random vector
describing the oriented fields of the subnets and $x_{i,\kappa}(\tb)$ 
the embedding strength following from the optimal learning strategy
for the subnet $i$ (without taking into account {\it response} $g$), then 
\beqn
1 \: = \: \alpha_i \int {\cal D} \tb \: [ \theta(x_\kappa(\tb)) + 
\kappa  x_{i,\kappa}(\tb) ]\quad .
\label{BooleCav}
\eeqn
  Here the different desired outputs, analogous to
the $(+)$ and $(-)$-patterns in case of the AND-machine, have to be
taken into account according to their respective probabilities. The
optimal storage capacity is obtained, if one of the subnets has
reached its capacity limit. If the output value follows from a Boolean
function, for which all the neurons in the intermediate layer are
equivalent, then all the $\alpha_i$ are identical. The  formula
analogous to (\ref{RSANDEnd}), giving an upper estimate for the
storage  capacity, was for
$\kappa =0$ already determined by Engel {\it et al.},
 \cite{EngelMehrschicht},
and was described in a more abstract way
 in \cite{GriniaGross,Griniasty}. 
 In a formulation  similar to  (\ref{BooleCav}), this upper bound is
determined from
 \beqn 1 \: = \: \alpha_i
\int {\cal D} \tb \: x_{i,\kappa}^2(\tb) \quad .  \eeqn

In contrast, the results for $\al_c$ obtained from
eqn.~(\ref{BooleCav}) with the cavity method are lower
estimates. For the {\it committee-machine}, because of
$\alpha_{\hbox{\tiny Cav}}<2$, they violate the {\it lower bound} of
Mitchison and
 Durbin \cite{Mitchison}, $\alpha>2$,
and are therefore without interest.

More interesting would be a RSB calculation as suggested at the
 end of the section on the simple perceptron for
 $\al > \al_c$, since then relevant estimates
from below for the true capacity for this model could be derived.

\subsection{The  AND-machine with overlapping receptive fields}
\label{ANDvollab}

Also the fully connected AND machine can be treated with our cavity method.
This correponds to a machine as in fig.\ 5, but with identical patterns
presented to both subnets. Thus the index $i=1,2$ of the description
of the patterns $\{\xi_k^\mu\}$, e.g.~in (\ref{JANDKopplungen}), is now dummy,
 and in particular it is
$\stackrel{\leftrightarrow}{B}_1 = \stackrel{\leftrightarrow}{B}_2
=: \stackrel{\leftrightarrow}{B}$.
A very important parameter for the fully connected AND machine is the
overlap $R \, := \, \sum_k J_{1k} J_{2k}$ of the two subnets.

Since also for this case one expects RSB, we can no longer expect 
 that $R$ agrees for different approaches, as it happened
with the generalization problem treated in \cite{Gerl3}.
Therefore we can no longer simply compare with the RS results of
\cite{GriniaGross}.

For the  local  fields $t_1$ and $t_2$ one gets for given $R$
\beqa
P_R(t_1,t_2) &=& \frac{1}{2\pi \: \sqrt{1-R^2}} \:
\exp \left( -\frac{t_1^2-2Rt_1t_2 + t_2^2}{2(1-R^2)} \right ) \quad .
\eeqa    
 However, the probability density of the field $t_1$ of a $(+)$-pattern needing
explicit embedding is not influenced, since again
\beqn
\int\limits_{-\infty}^{\infty} \hbox{d}t_2 P_R(t_1,t_2) \; = \;
\frac{1}{\sqrt{2\pi}}  e^{-t_1^2/2}   \quad .
\eeqn

For the $(-)$-patterns the situation is more complicated. 
Negative correlations of the subnets simplify the storing of these patterns,
whereas positive correlations  enhance the probability that 
a $(-)$-pattern, i.e.~one which is already wrongly classified by one
of the subnets, cannot be stored automatically by the other one,
i.e.~that it must be embedded explicitly.
 In the limit $b\to -1$, i.e.~when exclusively
 $(-)$-patterns must be embedded, 
 $\underline{J}_1 = - \underline{J}_2$ (with $R=-1$), for arbitrary
couplings
 $\underline{J}_1 $, is a general solution of the problem
in the limit $\alpha \to \infty$. 
 
For arbitrary $b$, the field of a $(-)$-pattern before learning is
\beqa
P_R(t) &=&  2 P_R(t_1=t, t_2<t) \; = \; 
\int\limits_{-\infty}^t \hbox{d}t_2 \,P_R(t,t_2) \nonumber\\
&=&
\frac{2}{\sqrt{2\pi}} \:e^{-t^2/2} \: \Phi \left ( \frac{t\,(1-R)}{\sqrt{1-R^2}}
\right )  \quad , 
\eeqa
and one obtains  (\ref{PvontNRF}) as special case for $R=0$. Again, a single 
groundstate is assumed, and again the arguments follow from
eqn.~(\ref{AntAND}) ff.

For the  {\it response} $g$ and the capacity $\alpha$ we get
\beqa
g & =&  -\alpha_+ \int\limits_{-\infty}^\kappa {\cal D} t
 \: - \: \alpha_- \int\limits_{-\infty}^\kappa {\cal D} t
\: \Phi \left ( \frac{t\,(1-R)}{\sqrt{1-R^2}}
\right ) \quad ,  \\
1 &=&  \alpha \int_{-\infty}^\kappa {\cal D}t
\left(
\frac{\raisebox{-0.03cm}{\hbox{\footnotesize 1 + {\it b}}} } 
 {\raisebox{0.03cm}{\hbox{\footnotesize 2}} }   \: +\: 
\frac{\raisebox{-0.03cm}{\hbox{\footnotesize 1 -- {\it b}}} } 
 {\raisebox{0.03cm}{\hbox{\footnotesize 2}} } 
\,\Phi \left ( \frac{t\,(1-R)}{\sqrt{1-R^2}}
\right )  \right) \quad .
\label{ANDvollkap}
\eeqa
Similarly as for the generalization problem in section 2.3 
of \cite{Gerl3},
only that solution $\alpha(R)$ is relevant, for which $R$ is reproduced
selfconsistently. For normalized couplings one obtains
\beqa
R & = &\sum_k J_{1k} J_{2k}  \:=\:
 \sum_k \sum_{\mu \nu} x_1^\mu \zeta^\mu \xi_k^\mu
x_2^\nu \zeta^\nu \xi_k^\nu  
= \sum_{\mu \nu} x_1^\mu B^{\mu \nu} x_2^\nu 
=  \sum_\mu x_1^\mu E_2^\mu \nonumber \\
&=&  \alpha_- \int\limits_{-\infty}^\kappa \hbox{d}t_1
\int\limits_{-\infty}^{t_1}  \hbox{d}t_2 \: P_R(t_1,t_2) \,\frac{\kappa -t_1}{1+g}
\: t_2  \nonumber      \\  && + \;
\alpha_+ \int\limits_{-\infty}^\kappa \hbox{d}t_1
\int\limits_{-\infty}^\infty  \hbox{d}t_2 
\: P_R(t_1,t_2) \,\frac{\kappa - t_1}{1+g} \:
\max \{ \kappa, t_2 \}  \quad . \label{ANDvollRkons}
\eeqa

In eqn~(\ref{ANDvollRkons}) the above-mentioned strategy for
 the storing of patterns was used:
For the $(-)$-patterns the subnet with the smaller oriented field,
here $t_2$,
 is unchanged, whereas  the $(+)$-patterns are explicitly embedded in
both
subnets $i$, if the field $t_i$ is $< \kappa$.
Again one multiplies with (1+$g$) to simplify 
(\ref{ANDvollRkons}).

For $\kappa=0$ it is again $g=-1$, and therefore one obtains $R$ from
\beqa
0& =&  \frac{1-b}{2} \int\limits_{-\infty}^0 \hbox{d}t_1
\int\limits_{-\infty}^{t_1}  \hbox{d}t_2 \: P_R(t_1,t_2) \,(-t_1)
\: t_2 \nonumber \\
&& + \; \frac{1+b}{2}\int\limits_{-\infty}^0 \hbox{d}t_1
\int\limits_0^\infty  \hbox{d}t_2 
\: P_R(t_1,t_2) \,( - t_1) \:
t_2  \quad .
\label{Rauskapnull}
\eeqa

The result for  $R(b)$ is presented in Fig.~7. One should note that it deviates
from the result obtained by Griniasty and Grossman in  \cite{GriniaGross}.
From a comparison with Fig.~3a in \cite{GriniaGross} one finds
that our value for $R$ is systematically larger, which - as mentioned
above -  has a negative effect on the storage capacity.

In Fig.~8 
 the capacity of the fully connected AND-machine
is presented for the case that the capacity in (\ref{ANDvollkap}) is
calculated for $\kappa=0$ with the  $R$ determined from
(\ref{Rauskapnull}).
For  $b=0$ we get $R=0.217$ and $\alpha = 3.113$, as opposed to 
 $R=0$ and $\alpha=3.512$ obtained by Griniasty and Grossman,
\cite{GriniaGross}. Again we expect that the
RS-approximation gives the better estimate. However, our estimate
for $b \lsim -0.65$ is above the {\it lower bound}
(\ref{Wendbed}), as it should, and in the limit $b \to -1$ one gets
 $\alpha_{\hbox{\scriptsize Cfull}} \to \infty$.

Apart from the fact that different capacities $\al_c$ are obtained
with the replica and the
cavity approach, there is a second hint on RSB:
The replica calculation with the  RS ansatz yields
an overlap $R$ between the subnets, which is not reproduced by our
``optimal'' learning algorithm.
 For the generalization problem, see \cite{Gerl3}, the
perfect agreement of the results obtained with the two approaches was
a clear hint on the correctness of the solution, and in particular on
the correctness of RS.
 In the present case, however, one finds that the RS solution
apparently yields
the optimal capacity only, when an additional component to the
coupling vector $\underline{J}$ is
introduced, by which the overlap $R$ between the subnets
is reduced, but the embedding of the patterns is actually weakened.

In fact, in \cite{Griniasty}, Griniasty suggests a non-trivial
training process for the fully connected AND-machine, by which
patterns in the non-affected subnet are unlearned, influencing in this
way the correlation $R$. According to our considerations, however,
this is not optimal, since after deleting the additional component one
can enhance the stability or store additional patterns, using the same
distribution of tasks with respect to the different patterns.
Thus it is not astonishing that Griniasty \cite{Griniasty} with his
training process obtains a smaller capacity ($\al_c=3.0$) as with a
stochastic algorithm, which leads to $\al_c=3.3$, \cite{GriniaGross}.
This discrepancy, which demands an additional component to the coupling
vector, by which the subnets are decorrelated, should
become smaller by a RSB calculation.

\vspace{1.5cm}

\section{Discussion} As we have stated already in [1], our cavity
 approach is usually technically simpler than a replica calculation.
Moreover, it gives the exact result as long as calculations within the replica
approach under the assumption of Replica Symmetry (RS) are
correct. Here we have shown additionally that for models with Replica-Symmetry
Breaking (RSB), e.g.~the simple perceptron above $\al_c$, one gets
different results, as one should, with our ''Kuhn-Tucker cavity
method'' and the RS
approximation: This would not be the case e.g.~with
 the different cavity approximation of
Griniasty \cite{Griniasty} or Wong \cite{Wong}, since
their methods are always equivalent to the replica calculation in RS
approximation. However, although our cavity theory ''indicates the
necessity of RSB'', if RS does not suffice, it is
 still far from being exact, since the combinatorial
 explosion of distributing the set of patterns into ''good''
patterns, which are stored, and ''bad ones'', which are not, is not 
considered.

 From the results of the present paper
 it can be seen in detail that for cases with RSB,
 Griniasty's ''RS-cavity theory'' approach usually yields
too optimistic estimates, whereas in the same cases,
 our  Kuhn-Tucker
cavity method is apparently ''too pessimistic'' in the error rates.  However,
 the limiting negative field value, below
which patterns are no longer learnable, (e.g.~$t\approx -0.5$ in
Fig.~3), is well approximated by our theory, and as shown above, the theory
also gives good results, if one starts with the field-distribution
obtained  by a 1-step-RSB calculation.

Moreover, the RS approximation follows for our models, except of the
last-mentioned case of the fully connected AND machine, always from
the ''formally crude'', but consistent approximation
 of vanishing response-factor $g$ and  vanishing off-diagonal 
elements $B^{\mu \nu}$, as stated already by Griniasty,
 \cite{Griniasty}.
For this fact we do not yet have a deeper understanding.

With our Kuhn-Tucker cavity-approach, we follow the embedding
of a new pattern in detail: The newly added pattern is 
 embedded by one single AdaTron step
with an enhanced implementation strength
$(1-\widetilde E^0)(1+g)^{-1}$, where the enhancement factor $-g$ is
given by eqn.~(\ref{gBeginn}).  At the same time, the already 
embedded patterns get specific corrections $\delta x^\mu = 
{\cal O}(1/\sqrt{N})$
of their implementation strengths. In the Appendix we show that for
$N\to\infty$ there are no further corrections for $g$ necessary, 
which would go beyond the one-step procedure. At the same time,
we have gained in this way knowledge of the actual distribution
of the embedding strengths.

Another important point of our cavity method is the demand that the
constraints, which the solutions impose on the couplings, are actually
realizable, which means that no more degrees of freedom are fixed than
are available with the given couplings.  For the fully connected
AND-machine an additional postulate is that the correlation of the
subnets is self-consistently reproduced by the embedding
strengths. Thus we can interprete our result as an estimate adapted to
our training algorithm.

Although replica calculations in RS approximation do not at all take
care of such details, we have found that they yield good estimates for
the storage capacity of the simple perceptron above $\al_c$, and
probably also for the AND machine. In this respect, the virtue of our
approach is based on two facts:

At first, it yields an independent
estimate, which seems to be a lower bound for $\al_c$ and an upper
bound for the error fraction $f(\alpha ,\kappa )$.

Second, our cavity theory visualizes the ''internal
stresses'' inherent in the RS approach and shows, which quantities and
order parameters depend most sensitively on the assumptions made.

Finally we repeat that our Kuhn-Tucker cavity approach, starting from
field-distri\-butions, which Majer and Engel, see \cite{MajerEngel},
obtained in 1-step RSB for
the perceptron above $\alpha_c(\kappa)$, 
shows that the 1-step RSB results are not yet exact, but
must already be very near to the truth. This  conclusion 
is supported by a recent 2-step RSB calculation, 
\cite{Sherrington}.

\subsection*{Acknowledgements}
The authors gratefully acknowledge helpful discussions 
with B.~Schottky, J.~Winkel, A.~Zippelius, and particularly 
 Frank R.~Bernhart, who supplied 
the elegant combinatoric arguments in the Appendix.

\section*{Appendix}
In this Appendix we show that for $N\to\infty$ our 1-step
 approximation for the reaction strength $g$ is exact:

In the derivation of our result for $g$
in eqn.~(\ref{gBeginn}), we had simply used $\delta x^\mu
=-B^{0\mu} x^0$ for the correction to the implementation
strengths of those patterns $\vec \xi ^\mu$,
which had already been embedded into the couplings
before the addition of the
test pattern $\vec\xi^0$. I.e.~we had neglected the (secondary)
mutual reaction of patterns with $\mu$ and $\nu \ge 1$ in
contrast to the (primary) response of the patterns $\vec \xi^\mu$ on the
the test
pattern.  To  be more thorough, 
let us thus try a correction term $y^\mu$ taking the 
 secondary and further reaction terms into account through $ 
\delta x^\mu =- B^{\mu 0}x^0 +y^\mu \,$.   Inserting this into
the equation  $ \delta E^\mu := B^{\mu 0}x^0 + \delta x^\mu +
\sum_{\nu (\ne \mu)=1}^p B^{\mu\nu}\delta x^\nu \,\solleq \, 0\,,$
we get iteratively
 \beqn \label{uniform00} \delta x^\mu =
x^0\cdot \left ( - B^{\mu 0}+\sum_{\nu (\ne \mu)=1}^p B^{\mu \nu}
B^{\nu 0}-\sum_{\nu (\ne \mu)=1}^p\sum_{\rho (\ne \nu)=1}^p
B^{\mu \nu}B^{\nu \rho}B^{\rho 0}\pm ... \right )\,. 
\eeqn 
Here, the 2nd and 3rd term on the r.h.s.~correspond to subsequent parallel
AdaTron iterations.
Indices corresponding to patterns,
 which are automatically implemented without explicit embedding, are
left out in the sums (which corresponds to 
$ \al\to\al_{\hbox{\scriptsize eff}} $ below).  
For the response $g$ we then have
\beqn 
 g \;=\;-\sum_{\mu}(B^{0 \mu})^2
+\sum_{\nu\ne\mu}B^{0 \nu}B^{\nu \mu}B^{\mu 0}
 - \sum_{\rho \ne\nu \ne\mu}
B^{0 \rho}B^{\rho \nu}B^{\nu \mu}B^{\mu 0} \pm \ldots\quad . 
\label{uniform}
\eeqn
We assume that there is no selection effect
among the correlation matrix elements, i.e. that if a pattern is
embedded explicitly this does not change the distribution of the
$B^{\mu \nu}$.

In eqn.~(\ref{uniform}) the first term on the r.h.s.~gives
\beqn
-\sum_\mu (B^{0 \mu})^2 \:=\:
-\frac{1}{N^2}\sum_{\mu} \sum_{i,j} \xi_i^0\xi_i^\mu\xi_j^\mu\xi_j^0
\:= \:    -\alpha_{\hbox{\scriptsize eff}} \;\: ,
\eeqn 
as used in eqn.~(\ref{Bmunuqu}).
The dominant contribution comes from $N$ terms such that $i=j$.
For the next term we have $N$ contributing terms $i=j=k$ plus remaining
terms represented by $\sum '$ below:
\beqa
\sum_{\nu\ne\mu}B^{0 \nu}B^{\nu \mu}B^{\mu 0}
&=& \frac{1}{N^3} \sum_{\nu\ne\mu}\sum_{i,j,k}  
\xi_i^0\xi_i^\mu\xi_j^\mu\xi_j^\nu\xi_k^\nu \xi_k^0 \\
&=&
\frac{p_{\hbox{\scriptsize eff}} (p_{\hbox{\scriptsize eff}}-1)N}
{N^3} + \frac{1}{N^3}\sum_{\nu\ne\mu} \sum_{i,j,k}{}'\,\, 
\xi_i^0\xi_i^\mu\xi_j^\mu\xi_j^\nu\xi_k^\nu \xi_k^0 \nonumber\\
&=& \alpha_{\hbox{\scriptsize eff}}^2 + {\cal O}(1/\sqrt{N}) 
\label{alphaqu}\; .
\eeqa
From the last explicit summation in (\ref{uniform}) there are just two
non-zero terms, one for $i=j=k=l$, the other one
for $\mu=\rho$, $i=l$ and $j=k$, giving
\beqa \label{eqn55}
- \sum_{\rho \ne\nu \ne\mu}
B^{0 \rho}B^{\rho \nu}B^{\nu \mu}B^{\mu 0} &=& 
- \frac{1}{N^4}\sum_{\rho\ne\nu\ne\mu} \sum_{i,j,k,l} 
\xi_i^0\xi_i^\mu\xi_j^\mu\xi_j^\nu\xi_k^\nu \xi_k^\rho \xi_l^\rho
\xi_l^0 \\
&=&
- \frac{p^2_{\hbox{\scriptsize eff}} (p_{\hbox{\scriptsize eff}}-1)N}
{N^4} - \frac{p_{\hbox{\scriptsize eff}} (p_{\hbox{\scriptsize eff}}-1)N^4}
{N^4} + {\cal O}\big(\frac{1}{\sqrt{N}}\big)\nonumber
\\ 
&\simeq&
- \alpha_{\hbox{\scriptsize eff}}^3 -\alpha_{\hbox{\scriptsize eff}}^2
\; \quad .\label{eqn56}
\eeqa
Thus the term $\alpha_{\hbox{\scriptsize eff}}^2$ in (\ref{alphaqu})
is cancelled.

In the following we will see that the same happens for all powers
of $\alpha_{\hbox{\scriptsize eff}}$ that appear at some point in the
series.

First we observe, that the power of $\alpha_{\hbox{\scriptsize eff}}$
in a term is given by the number of different pattern indices
$\mu,\nu,\ldots$ we sum over.  As in the example above, eqs.~(\ref{eqn55}),
 (\ref{eqn56}), we can
eliminate pattern indices by summing over equal pairs.  Because of the
construction of the parallel AdaTron algorithm no next neighbours in a
sum (e.g.~$\mu ,\nu$ above) can be eliminated.  Next we observe, that
when we eliminate a pair of pattern indices (e.g.~$\mu = \rho$ above),
all neuron indices $i,j,\ldots$ in between have to be put equal
(e.g.~$j=k$ for the 2nd term mentioned in connection with
 eqn.~(\ref{eqn55})). 
Thus trying to "join" a pattern index
within a pair with a pattern index outside gives no contribution. 
In other words, once we have $\mu=\omega$ in the chain 
$\mu\nu\rho\sigma\omega\eta\theta$ putting $\rho=\theta$ 
does not make any sense
(see the case $\ldots x\ldots y \ldots x \ldots y
  \ldots $  below).

The problem of enumerating the number of different ways
of contributions that can appear can be solved with a
little help from combinatorics \cite{Bernhart}.
First we reformulate our problem:
We have a sequence of $n$ symbols $a,b,c,\ldots$, for
instance $[abcdba]$, which fulfill:
\begin{itemize}
\item The first occurence of every symbol must be in alphabetic order.
\item The same symbol cannot occur twice consecutively.
\item There is no subsequence $\ldots x\ldots y \ldots x \ldots y
  \ldots $ unless $x=y$.  \end{itemize} Here the symbols correspond to
our pattern indices $\mu$, $\nu$, $\rho$, $\ldots$.

There is exactly one way for all symbols to be different, 
and there are $(n-1)(n-2)/2$ ways for exactly one pair.
There are exactly two ways for 2 identities (counting the number
of ``='' signs needed to fix the pattern indices)
in a chain of 5 symbols, namely $[abaca]$ and $[abcba]$. 
There are 5 possible arrangements in a chain of 7 letters
which permit 3 eliminations of patterns indices:
$[abacada]$, $[abcbada]$, $[abacdca]$, $[abcdcba]$ and
$[abcbdba]$. In both cases no additional identity is possible,
and we see that the number $n$ of pattern indices has to
be larger than $2k+1$, with $k$ being the number of
identities.

The number $(n,k)$ of cases, where one has $n$ letters and $k$
identities, can be calculated by using a {\it generating function}.
For small values of $n$ and $k$ the numbers are shown in table 1.  Let
us define a function $y(w,x)$ of two real variables $w$ and $x$ given
by the power series \beqa \label{eqn57} y(w,x) &=& \sum_{n= 1}^\infty
\sum_{k=0}^{[(n-1)/2]} (n,k)\, w^k x^n \nonumber  \\
       &=& x + x^2 + (1+w)x^3 + (1+3w)x^4 + \ldots \quad .
\eeqa
We can obtain the defining equation for $y$ by recursion:
If we remove the first letter in the sequence,
the possibilities are that
\begin{itemize}
\item there is nothing left
\item the letter does not appear again, and there are no
      restrictions imposed on the rest of the sequence
\item the letter appears in the rest of the sequence, and
      because of the last condition above we now have two sequences
left (e.g.~$[bcb]$ and $[ada]$ in the example $[abcbada]$).
\end{itemize}
These contributions give the terms $x$, $xy$ and $wxy^2$ respectively.
Thus we have for the defining equation 
\beqa
y&=& x(1+y+wy^2) \quad .
\eeqa
Using the B\"urmann-Lagrange series \cite{Courant} 
for inversion on this equation tells us
that
\beqa
\sum_{k=0}^{[(n-1)/2]}\:(n,k) \: w^k &=& 
\frac{1}{n!} \frac{\partial^{n-1}}{\partial t^{n-1}} \:
            (1+t+wt^2)^n\vert_{t=0}\,\,. 
\eeqa
This can be written in a more convenient way:
\beqa
n\cdot (n,k) &=& \hbox{ the coefficient of } \; w^k t^{n-1} \hbox{ in }
 (1+t+wt^2)^n\,.
\eeqa
Thus we have
\beqa
((1+t)+wt^2)^n= \sum_{k=0}^n{n\choose k}\sum_{l=0}^l {n-k\choose l} w^k
  t^{2k+l}
\; ,
\eeqa
and putting $2k+l = n-1$ to get the coefficient of $w^k t^{n-1}$ we arrive at
\beqa
(n,k) = \frac{1}{n} \; { n \choose k} {{n-k} \choose {k+1}} \; .
\label{nkFormel}
\eeqa
If we put $w=1$, we sum the rows in table 1, which gives the defining
equation $y=x(1+y+y^2)$ for the Motzkin numbers 1 1 2 4 9 21 51 127$\ldots$.
The last numbers in lines $n=1,3,5,\ldots$ are the Catalan series.

We remember that $(n,k)$ is the coefficient of 
$\alpha_{\hbox{\scriptsize eff}}^{n-k}$ 
in the $n$-th iteration of
the AdaTron algorithm. 
 We want to prove that in the thermodynamic limit the first term
is already sufficient; thus
 we have to show that
$\sum_{k=0}^{n-1} (-1)^k (n+k,k) = 0$ for $n\ge 2$.
Substituting $w/x$ for $w$ in the defining equation moves
the column for each $k$ in the table up k steps from the beginning. 
After rearranging we have $y= x(1+y)/(1-wy)$.
To produce  an alternating sum we now use $w=-1$ and  arrive
at $y= x+ 0 + 0 +\ldots$, just as we wanted. 

However, after $n$ such iterations of our simple parallel
 AdaTron algorithm there
is still a ``tail'' with powers of $\alpha_{\hbox{\scriptsize eff}}$
ranging from $[(n-1)/2]$ to $n$, which have not yet been eliminated.
For $\alpha_{\hbox{\scriptsize eff}} \gsim 1/3$
this tail (and the results of the simple AdaTron algorithm)
 will oscillate with increasing amplitude.   
But if we now introduce as usual an overrelaxation parameter
$\gamma$ small enough, these oscillations are damped out
and the modified AdaTron algorithm  $\delta
x^\mu=\max{(-x^\mu,\gamma(1-E^\mu))}$ 
converges for all $\alpha_{\hbox{\scriptsize eff}} < 1 $ \cite{Anlauf}. 
At the same time we will see that our cavity response theory 
is correct already after the first step: 

Since we are not concerned with computational efficiency, we can
choose an {infinitesimally small} $\gamma$,
 $N^{-1/2} \ll \gamma \ll 1$, {\it after} the
first AdaTron step, to examine convergence of the above-mentioned tail.
The number $l$ of AdaTron steps  of course has to 
be increased in correspondence to the reduction of $\gamma$,
 so that the product $\gamma l$ remains finite.
For the first few steps of this modified AdaTron algorithm 
with overrelaxation $\gamma$ after the first step, the
response at pattern 0 then reads: 
\begin{eqnarray}
g_2 &=& 
-\alpha_{\hbox{\scriptsize eff}} + \gamma  \alpha_{\hbox{\scriptsize
      eff}}^2 \nonumber \\
g_3 &=& -\alpha_{\hbox{\scriptsize eff}} + 
(\gamma (1\!-\!\gamma) + \gamma)   \alpha_{\hbox{\scriptsize
      eff}}^2  - \gamma^2 (\alpha_{\hbox{\scriptsize eff}}^2 + 
\alpha_{\hbox{\scriptsize eff}}^3) \nonumber \\
g_4&=& -\alpha_{\hbox{\scriptsize eff}}  +
(\gamma (1\!-\!\gamma)^2 + \gamma(1\!-\!\gamma) + \gamma) 
\alpha_{\hbox{\scriptsize eff}}^2
- (2 \gamma^2(1\!-\!\gamma)+ \gamma^2) (\alpha_{\hbox{\scriptsize eff}}^2
+ \alpha_{\hbox{\scriptsize eff}}^3)+ \gamma^3
(3\alpha_{\hbox{\scriptsize eff}}^3 +\alpha_{\hbox{\scriptsize eff}}^4)
\nonumber\\
&\vdots& \nonumber\\
g_l &=& -\alpha_{\hbox{\scriptsize eff}} + \gamma \sum_{i=0}^{l-2}
(1-\gamma)^i \:\alpha_{\hbox{\scriptsize eff}}^2 - 
\gamma^2 \sum_{i=1}^{l-2} i(1-\gamma)^{i-1}
(\alpha_{\hbox{\scriptsize eff}}^2 +\alpha_{\hbox{\scriptsize eff}}^3)
+  \nonumber \\  && \qquad 
\gamma^3 \sum_{i=2}^{l-2} \frac{i(i-1)}{2} (1-\gamma)^{i-2}
(3\alpha_{\hbox{\scriptsize eff}}^3 +\alpha_{\hbox{\scriptsize eff}}^4
)  - \ldots \quad .\nonumber 
\end{eqnarray}
Summing the geometrical series and using $(1-\gamma)^l \simeq e^{-\gamma l}$
for $\gamma \to 0^+$ and $l \to \infty$, we get 
\begin{eqnarray}
g_l &\simeq& - \alpha_{\hbox{\scriptsize eff}} + 
(1- e^{-\gamma l}) \alpha_{\hbox{\scriptsize eff}}^2
- (1-e^{-\gamma l}(1 + \gamma l)) (\alpha_{\hbox{\scriptsize eff}}^2
+ \alpha_{\hbox{\scriptsize eff}}^3) \nonumber\\
&& + (1-e^{-\gamma l}\big(1 + \gamma l + (\gamma l)^2/2 )\big ) 
(3\alpha_{\hbox{\scriptsize
    eff}}^3+ \alpha_{\hbox{\scriptsize eff}}^4 ) - \ldots \nonumber\\
&=& -\alpha_{\hbox{\scriptsize eff}} +
\sum_{n=2}^\infty (-1)^n \Big( 1- e^{-\gamma l} 
\sum_{m=1}^{n-1} \frac{(\gamma l)^m}{m{\rm !}} \Big )
\sum_{k=0}^{[(n-1)/2]} (n,k) \alpha_{\hbox{\scriptsize eff}}^{n-k}  
\nonumber \\
&=& -\alpha_{\hbox{\scriptsize eff}} + e^{-\gamma l}
\sum_{n=2}^\infty
 \alpha_{\hbox{\scriptsize eff}}^n
\sum_{k=0}^{n-2} (-1)^k \frac{(2n-k-2)(\gamma l)^{2n-k-3}}
 { n(n-1)(n-k-2){\rm!}(n-k-1){\rm !}k{\rm !} }\,\,\,\,.
\label{alphafinform}
\end{eqnarray}
Numerically the sum in (\ref{alphafinform}) decays
to 0.  Convergence is faster as $\alpha_{\hbox{\scriptsize eff}}$ decreases, just as 
expected. For the critical value $\alpha_{\hbox{\scriptsize eff}}=1$,
where $g \to -1$ and $\alpha \to \alpha_c$, 
we can examine the convergence analytically:

Collecting the series in (\ref{alphafinform}) and expanding
$e^{-\gamma l}$ gives for $\alpha_{\hbox{\scriptsize eff}}=1$ 
\begin{eqnarray}
g_l +1 &\simeq& \sum_{n=1}^\infty
(-1)^{n-1} \frac{6(2n-1){\rm !}(\gamma l)^n} 
{((n-1){\rm !})^2/(n+2){\rm !}} \nonumber\\
&=& \gamma l \cdot _{\:\: 1 \!\!}F_1(3/2,4,-4\gamma l) \label{hypergeom}\\
&=& \frac{16 \gamma l}{\pi} \int_0^1 e^{-4\gamma l t} \sqrt{t}
\, (1-t)^{3/2} \, \hbox{d}t\quad .  \label{intrep}
\end{eqnarray}
$_1F_1$ in eqn.~(\ref{hypergeom}) is the confluent hypergeomtric function,
in eqn.~(\ref{intrep}) we introduce its integral representation.
We are interested in the behaviour for $\gamma l \to \infty$ 
(while $\gamma \ll 1$),
therefore we can replace the $(1-t)^{3/2}$ term in the integral by 1
and perform the integration analytically.
Thus the additional disturbance decays like
\begin{eqnarray}
\lim_{\gamma l \to \infty} g_l \;+1 
&=&   \frac{1}{\sqrt{\pi \gamma l}}\to 0 \; .
\end{eqnarray}

We now have shown, that even for the critical value 
$\alpha_{\hbox{\scriptsize eff}}=1$ the additional terms
decay  $\to 0$ like $1/{\sqrt{\pi\, \gamma l}}$; 
for $\alpha_{\hbox{\scriptsize eff}}<1$, which are the values we
are interested in, numerical examination universally shows 
even faster decay.      Therefore, for
$N\to \infty$, our 1-step-reaction approach for $g$, which is
completely in the spirit of Onsager's cavity approach, is correct.
This is of course also true for the multilayer perceptrons studied. 

A further study, \cite{Biehldoktor}, shows that the patterns which are
explicitly embedded by the AdaTron algorithm, have a slight negative
correlation. However this small selection effect ${\cal O}(1/N)$
contributes even less than the correction effects treated above.
The same is true for the selection effect introduced by the 
combinatorial explosion, which allows to look for an optimal
groundstate. At the level of the correlation matrix elements $B$,
this again results in an effect  ${\cal O}(1/N)$ and can be neglected.

\newpage

\centerline{\bf Figure Captions}
\vglue 0.5 truecm
Fig.~1: For two different error rates $f$ the 
storage capacity $\alpha$ is presented as a function of
the stability $\kappa$ 
for both estimates
(\ref{KTresult}) and (\ref{obRSausger}).

Fig.~2: The stability $\kappa$ is presented as a function of the
error rate $f$ for
 $\alpha=2.0$, 0.8 and 0.4 . 
Results are shown for the 
cavity-method and the replica calculation
in RS and 1-step-RSB approximation. 

Fig.~3: The probability density $P(t)$ of the local fields $t$ before learning
is presented as a function of $t$ at
 $\alpha=1$ and $\kappa=1$. Again results are shown for
the cavity-method and the replica calculation
in RS and 1-step-RSB approximation. 
The distribution after training is given by pushing the
middle segment to a
 $\delta$-Peak at $\kappa=1$ in every case.
The cavity-method yields an error rate $f=0.29092$, the replica method
gives $f=0.13073$ in RS and $f=0.13576$ in 1-step-RSB approximation.
The fraction of explicitly learned patterns is 0.55042 for the
cavity approximation and 0.71062 and 0.66745 respectively for the
replica calculations.

Fig.~4: The error rate $f$ is presented as a function of
$\alpha$ for $\kappa=1.25$, $0.5$ and $0$.
The results are given for the cavity method, the RS and 1-step-RSB
approximation as well as the result
 $f_{\hbox{\tiny KT}}$, which is calculated by putting together the
 Kuhn-Tucker conditions and the local fields of the RSB-solution.  The
best estimate is very likely between these RSB-graphs,
i.e.~$f_{\hbox{\tiny
RSB}}$ and $f_{\hbox{\tiny KT}}$, and probably closer to $f_{\hbox{\tiny
RSB}}$. 

Fig.~5: The AND-machine
with tree structure (NRF). The weights between the intermediate layer and
the output neuron are fixed. A threshold between 0 and 1 at the
output-neuron takes care that there only is a positive output if both
intermediate neurons are positive.
 
Fig.~6:
The storage capacity of the AND-machine is presented as a function of
$\kappa$ for $b=0$. The results for the cavity method and for the
replica calculation are given. The storage capacity
 $\alpha_{\hbox{\scriptsize full}}$
of the fully connected AND-machine, see \ref{ANDvollab}, 
and 
$\alpha_{\hbox{\scriptsize Perc}}$
of the simple perceptron are presented for comparison

Fig.~7:
The overlap $R$ of both subnetworks of the fully connected AND-machine
is presented as a function of the 
 bias parameters $b$ from (\ref{ANDbias}). 
The result for the cavity method gives a higher correlation of the
subnetworks compared to
 {\it Fig.} 3.a in \cite{GriniaGross}.
We have $R=0$ at $b= -1/3$, which means $\alpha_-= 2\alpha_+$,
in contrast to $R=0$ at $b=0$ in \cite{GriniaGross}.

Fig.~8:
The maximal storage capacity of the AND-machine is presented
as a function of the
bias parameter $b$ from (\ref{ANDbias}). 
The result of the cavity method for the tree structure is compared 
to the RS-result from \cite{GriniaGross} and the cavity result
for the fully connected AND-machine.
The largest deviations occur for small values of $b$.
For the fully connected AND-machine the storage capacity is again
smaller than with the replica calculation
(see {\it Fig.} 3.b in \cite{GriniaGross}).
We have $\alpha_{\hbox{\scriptsize C-full}} = 
\alpha_{\hbox{\scriptsize C-tree}}$
at $b=-1/3$, when $R=0$. 
For $b \to -1$ we have $\alpha_{\hbox{\scriptsize C-full}}
\to \infty$ as in \cite{GriniaGross}.

{{\bf Table 1:} The numbers $(n,k)$ of possible arrangements of $n$ letters with
$k$ identities as described in the text. At the same time, these numbers are
the coefficents of $\alpha_{\hbox{\scriptsize eff}}^{n-k}$ in
the $n$-th term on the r.h.s.~of eqn.(\ref{uniform}). 
In the main text
it is shown, that the alternate sum along the diagonals --
connected as a guide to the eye -- disappears:
$\sum_l (-1)^l(n+l,l)=0$. E.g.~$1-3+2=0$, $1-6+10-5=0$ etc..
}

\newpage
\input epsf
\begin{figure}[htb]
\epsfxsize=15cm
\epsfbox{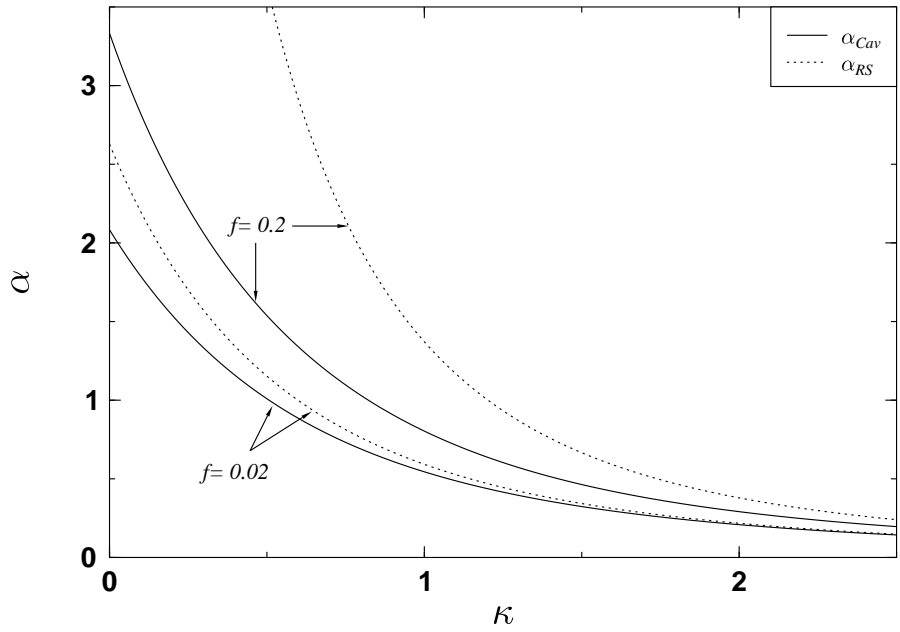}
\vglue 1.5 truecm
\caption[Speicherkapazitaet bei vorgegebener Fehlerrate]
{
For two different error rates $f$ the 
storage capacity $\alpha$ is presented as a function of
the stability $\kappa$ 
for both estimates
(\ref{KTresult}) and (\ref{obRSausger}).
}
\label{SpeichvonFehler}
\end{figure}

\begin{figure}[htb]
\epsfxsize=15cm
\epsfbox{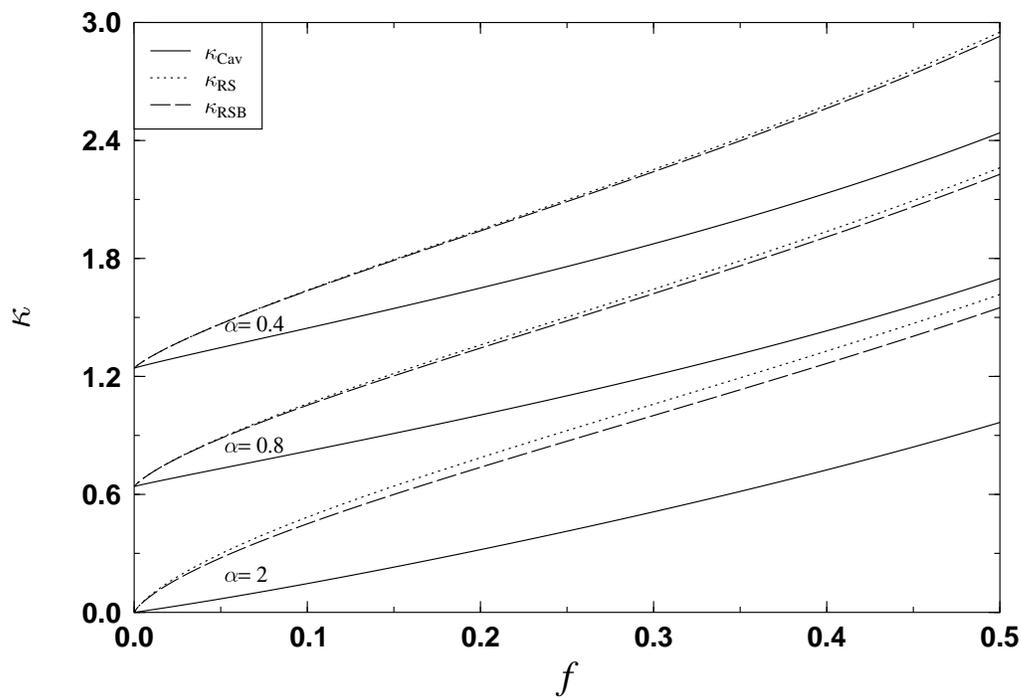}
\caption[Stabilitaetserhoehung durch Fehler]
{
The stability $\kappa$ is presented as a function of the
error rate $f$ for
 $\alpha=2.0$, 0.8 and 0.4. 
Results are shown for the 
cavity-method and the replica calculation
in RS and 1-step-RSB approximation. 
}
\end{figure}

\begin{figure}[htb]
\epsfxsize=15cm
\epsfbox{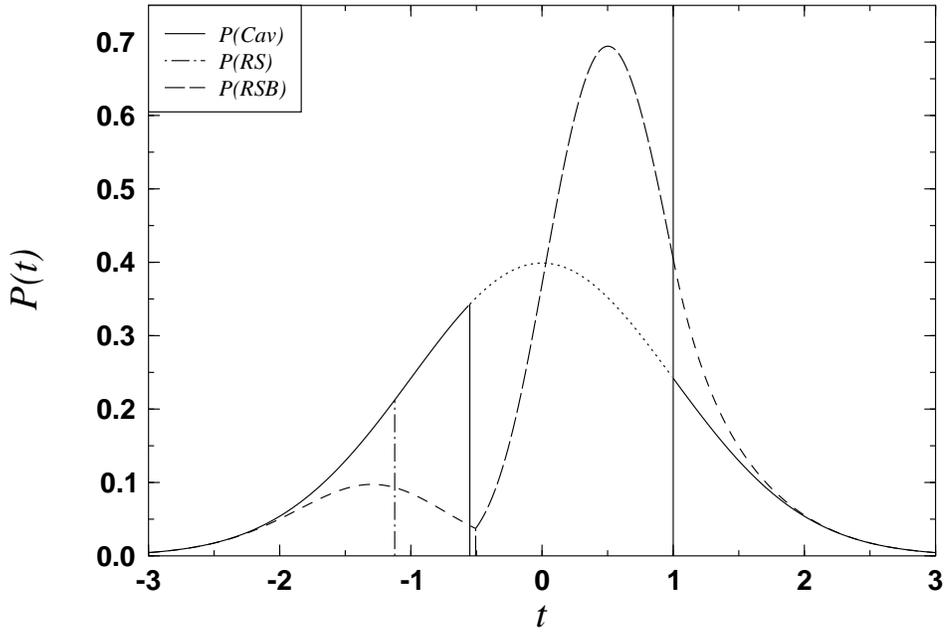}
\caption[Feldverteilung mit RSB]
{
The probability density $P(t)$ of the local fields $t$ before learning
is presented as a function of $t$ at
 $\alpha=1$ and $\kappa=1$. Again results are shown for
the cavity-method and the replica calculation
in RS and 1-step-RSB approximation. 
The distribution after training is given by pushing the
middle segment to a
 $\delta$-Peak at $\kappa=1$ in every case.
The cavity-method yields an error rate $f=0.29092$, the replica method
gives $f=0.13073$ in RS and $f=0.13576$ in 1-step-RSB approximation.
The fraction of explicitly learned patterns is 0.55042 for the
cavity approximation and 0.71062 and 0.66745 respectively for the
replica calculations.
}
\end{figure}

\begin{figure}[htb]
\epsfxsize=15cm
\epsfbox{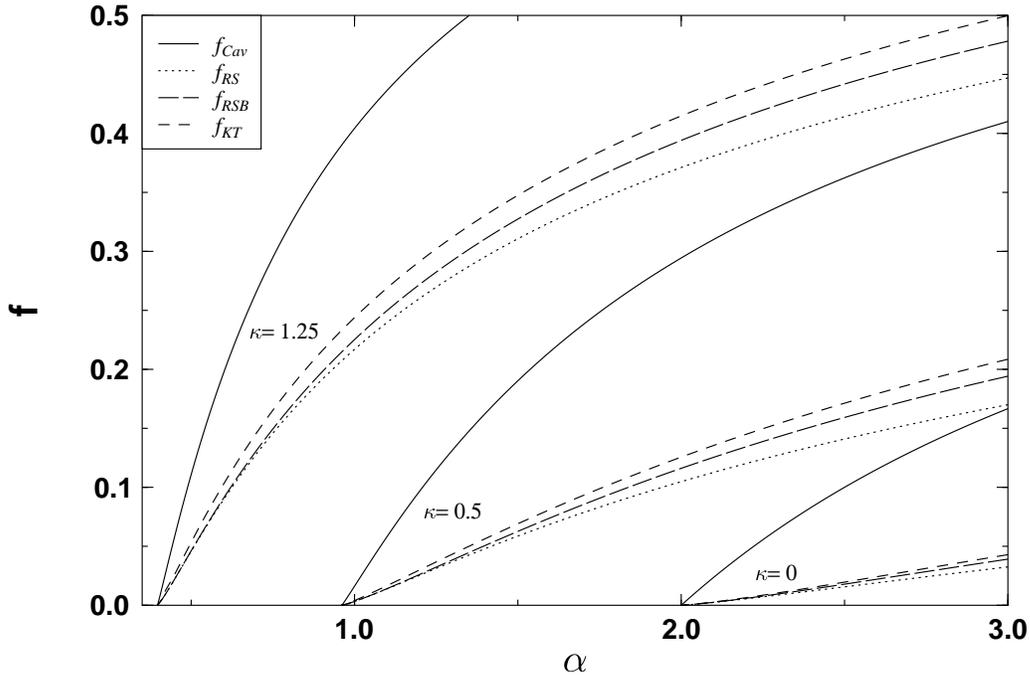}
\caption[Fehler von alpha mit allen Kurven]
{
The error rate $f$ is presented as a function of
$\alpha$ for $\kappa=1.25$, $0.5$ and $0$.
The results are given for the cavity method, the RS and 1-step-RSB
approximation as well as the result
 $f_{\hbox{\tiny KT}}$, which is calculated by putting together the
 Kuhn-Tucker conditions and the local fields of the RSB-solution.  The
best estimate is very likely between these RSB-graphs,
i.e.~$f_{\hbox{\tiny
RSB}}$ and $f_{\hbox{\tiny KT}}$, and probably closer to $f_{\hbox{\tiny
RSB}}$.  } \label{fRSBundKuhn} 
\end{figure} 

\begin{figure}[htb]
\epsfxsize=15cm 
\epsfbox{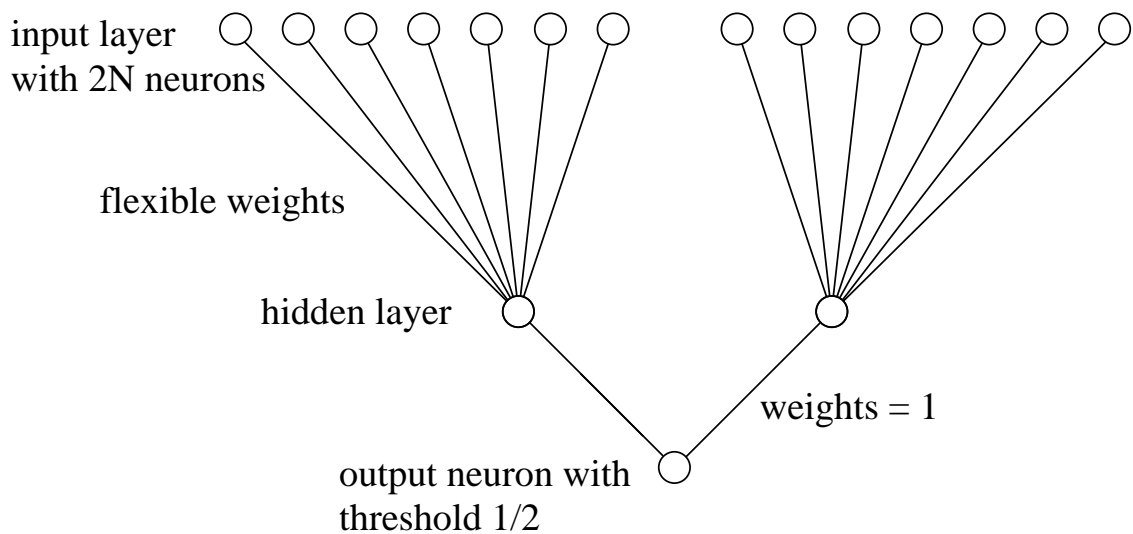}
\caption[Darstellung der AND-machine] 
{
The AND-machine
with tree structure (NRF). The weights between the intermediate layer and
the output neuron are fixed. A threshold between 0 and 1 at the
output-neuron takes care that there only is a positive output if both
intermediate neurons are positive.
  } \label{ANDdarst}
 \end{figure}

\begin{figure}[htb] \epsfxsize=15cm 
\epsfbox{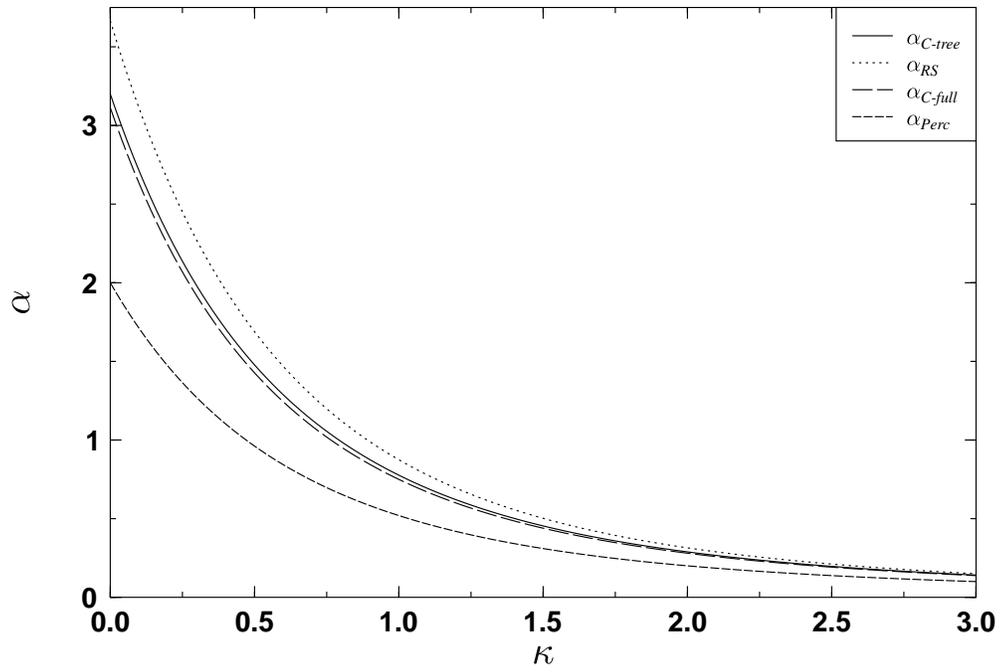}
\caption[Kapazitaet der AND-machine aus Stabilitaet] 
{
The storage capacity of the AND-machine is presented as a function of
$\kappa$ for $b=0$. The results for the cavity method and for the
replica calculation are given. The storage capacity
 $\alpha_{\hbox{\scriptsize full}}$
of the fully connected AND-machine, see \ref{ANDvollab}, 
and 
$\alpha_{\hbox{\scriptsize Perc}}$
of the simple perceptron are presented for comparison
}
\label{ANDKapStab}
\end{figure}

\begin{figure}[htb]
\epsfxsize=15cm
\epsfbox{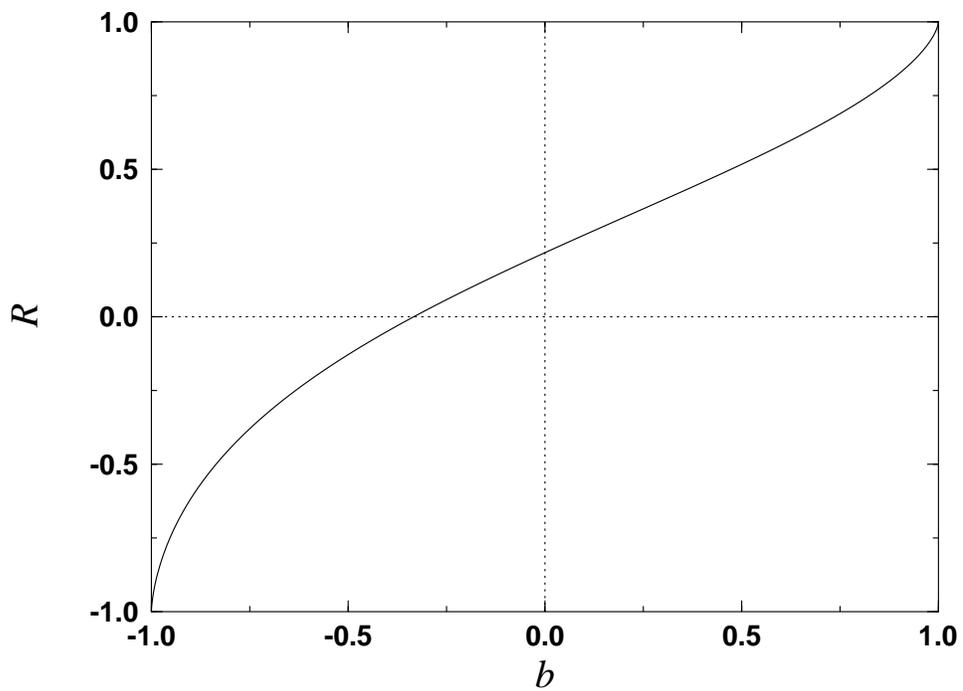}
\caption[Korrelation der Subnetzwerke abhaengig von b]
{
The overlap $R$ of both subnetworks of the fully connected AND-machine
is presented as a function of the 
 bias parameters $b$ from (\ref{ANDbias}). 
The result for the cavity method gives a higher correlation of the
subnetworks compared to
 {\it Fig.} 3.a in \cite{GriniaGross}.
We have $R=0$ at $b= -1/3$, which means $\alpha_-= 2\alpha_+$,
in contrast to $R=0$ at $b=0$ in \cite{GriniaGross}.
}
\label{Rvonb}

\end{figure}
\begin{figure}[htb]\epsfxsize=15cm
\epsfbox{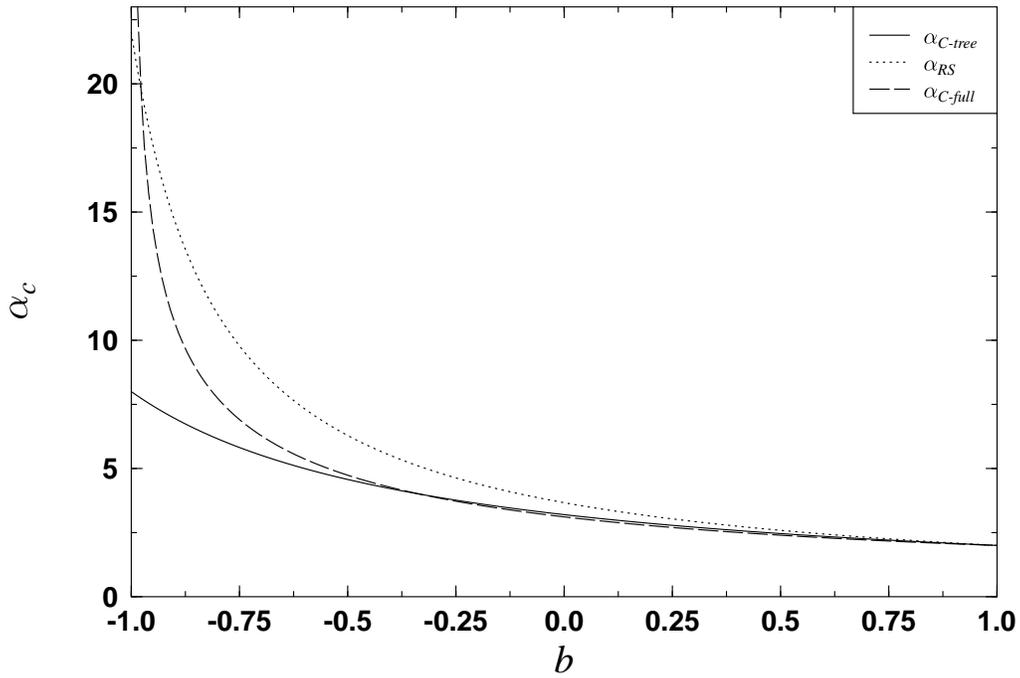}
\caption[Kapazit"at der AND-machine abhaengig von b]
{
The maximal storage capacity of the AND-machine is presented
as a function of the
bias parameter $b$ from (\ref{ANDbias}). 
The result of the cavity method for the tree structure is compared 
to the RS-result from \cite{GriniaGross} and the cavity result
for the fully connected AND-machine.
The largest deviations occur for small values of $b$.
For the fully connected AND-machine the storage capacity is again
smaller than with the replica calculation
(see {\it Fig.} 3.b in \cite{GriniaGross}).
We have $\alpha_{\hbox{\scriptsize C-full}} = 
\alpha_{\hbox{\scriptsize C-tree}}$
at $b=-1/3$, when $R=0$. 
For $b \to -1$ we have $\alpha_{\hbox{\scriptsize C-full}}
\to \infty$ as in \cite{GriniaGross}.
}
\label{ANDkapb}
\end{figure}

\begin{figure}
\epsfbox{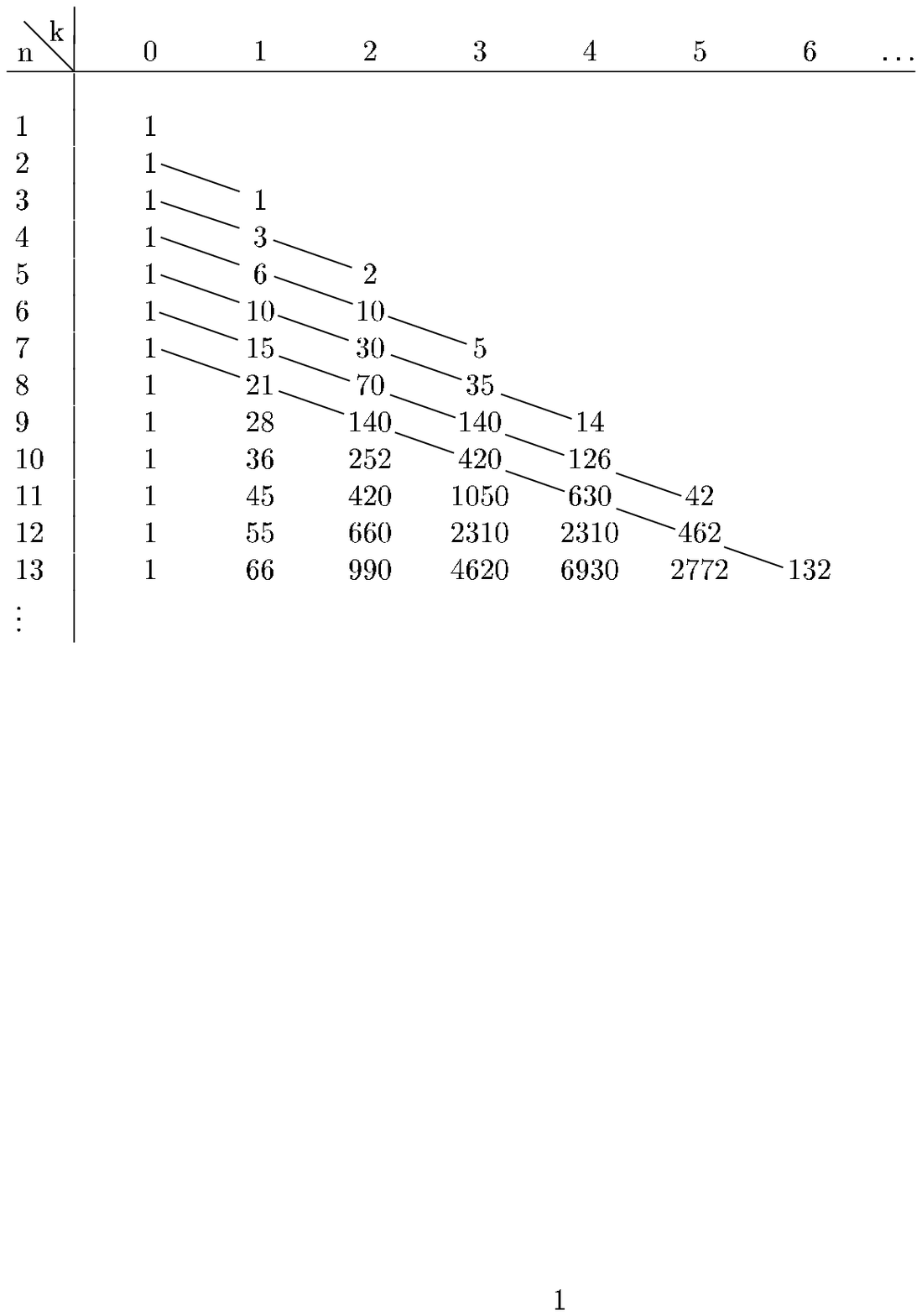}
\caption[Tabelle der Koeffizienten]
{{\bf Table 1:} The numbers $(n,k)$ of possible arrangements of $n$ letters with
$k$ identities as described in the text. At the same time, these numbers are
the coefficents of $\alpha_{\hbox{\scriptsize eff}}^{n-k}$ in
the $n$th term on the r.h.s.~of eqn.(\ref{uniform}). In the main text
it is shown, that the alternate sum along the diagonals --
connected as a guide to the eye -- disappears:
$\sum_l (-1)^l(n+l,l)=0$. E.g.~$1-3+2=0$, $1-6+10-5=0$ etc..
}
\label{table1}
\end{figure}


\begin{thebibliography}{99}
\bibitem{Gerl1} Gerl, F., Krey, U.: J. Phys. {\bf A 27}, 7353 (1994) 
\bibitem{Gerl3} Gerl, F., Krey, U.: J.~Phys.~{\bf A 28}, 6501 (1995) 
\bibitem{Winkel} Winkel, J.\'O., Gerl, F., Krey, U.: Z.~Physik {\bf B 100},
 149 (1996).
\bibitem{Mezard1} M\'ezard, M., J.~Phys.~{\bf A 22}, 2181 (1989) 
\bibitem{Kinzel} Kinzel, W., Opper, M., {\it Dynamics of Learning}, 
in: Physics of Neural Networks, (L. van Hemmen, E.~Domany and
K.~Schulten, eds.), Berlin, Heidelberg, New York, Springer 1991, p.149
\bibitem{Griniasty} Griniasty, M.: Phys.~Rev. {\bf E 47}, 4496 (1993) 
\bibitem{Wong} Wong, K.Y.M.: Europhys. Lett., {\bf 30} (4), 245 (1995)
\bibitem{GardnerDerrida} Gardner, E., Derrida, B.: J.~Phys.~{\bf A
21}, 271 (1988) 
\bibitem{Bouten} Bouten, M.: J.~Phys.~{\bf A 27}, 66021 (1994) 
\bibitem{GriniaGutfreund} Griniasty, M., Gutfreund, H.: J.~Phys.~{\bf
24}, 715 (1990) 
\bibitem{MajerEngel} Majer, P., Engel, E., Zippelius, A.:
J.~Phys.~{\bf A 26}, 7405 (1993) 
 \bibitem{Anlauf}  Anlauf, J.K., Biehl, M.: Europhys. Lett.
  {\bf 10}, 687 (1989) 
 \bibitem{Opper} Opper, M.: Phys. Rev. {\bf A38}, 3824 (1988) 
\bibitem{FontanariTheumann} Fontanari, J.F., Theumann, W.K.,
J.~Phys.~{bf A 26}, L1233 (1993)
\bibitem{Sherrington} Whyte, W., Sherrington, D.: J.~Phys.~{\bf A 29},
3063 (1996) 
\bibitem{Mezard} M\'ezard, M., Parisi, G., Virasoro, M.A.:
{\it Spin glass theory and beyond}, World Scientific, Singapore 1987
\bibitem{GriniaGross} Griniasty,M., Grossman, T., Phys.~Rev.~{\bf A 45}
8924 (1992)
\bibitem{MinskyPapert} Minsky, M.L., Papert, S.A., {\it  Perceptrons}, MIT
Press, Cambridge MA., 1969.
\bibitem{Mitchison} Mitchison, G.J, Durbin, R.M., Biol.~Cybernetics
{\bf 60} 345 (1989)
\bibitem{BarkaiHansel} Barkai, E., Hansel, D., Sompolinsky, H.,
Phys.~Rev. {\bf A 45} 4146 (1992)
\bibitem{EngelMehrschicht} Engel, A., K\"ohler, H.M., Tschepke, F.,
Vollmayr H., Zippelius, A., Phys.~Rev. {\bf A 45} 7590 (1992)
\bibitem{Wendemuth} Wendemuth, A: Int.~J.~Neural Systems {\bf 5},
 217 (1994)
\bibitem{Gardner}  Gardner, E.: Europhys. Lett. {\bf 4}, 481 (1987)
\bibitem{BiehlAnlaufKinzel} Biehl, M., Anlauf, J.k., Kinzel, W., in:
{\it Neurodynamics}, Eds. F. Pasemann and H.D. Doebner,
World Scientific, Singapore 1991  
\bibitem{Bernhart} Bernhart, F.R.: private communication
\bibitem{Courant} Hurwitz, A., Courant, R., {\it Allgemeine
Funktionentheorie und Elliptische Funktionen}, 4th ed.,  
Springer Verlag, Berlin,1964, p.~138
\bibitem{Biehldoktor} Biehl, M., PhD thesis, University of Giessen, 1992
\end{thebibliography}
\end{document}